# Near-Room-Temperature Antiferromagnetic Ordering in the Quadruple Perovskite $Sr_4NaRu_3O_{12}$

Subham Naik[a], Biswajit Singh[a], Hiranmayee Senapati[a], Akshay K. U.[b], Ramesh C. Nath[b], Soumyojit Chatterjee[a], Rahul Sharma[a], Thomas Doert[c], Walter Schnelle[d], Manfred Reehuis[e], Thomas C. Hansen[f], Michael Ruck[c,g], and Gohil S. Thakur[a,c,g*]

[a]*Department of Chemical Sciences, Indian Institute of Science Education and Research, Berhampur, 760010, India*

[b]*School of Physics, Indian Institute of Science Education and Research, Thiruvananthapuram, 695551, India*

[c]*Faculty of Chemistry and Food Chemistry, Technical University Dresden, 01069 Dresden, Germany*

[d]*Max-Planck Institute for Chemical Physics of Solids, 01187, Dresden, Germany*

[e]*Helmholtz Center Berlin for Materials and Energy, Hahn-Meitner-Platz 1, 14109 Berlin, Germany*

[f]*Institut Laue-Langevin, 71 Avenue des Martyrs, 38000 Grenoble, France*

[g]*Würzburg-Dresden Cluster of Excellence, ctd.qmat, 01069 Dresden, Germany*

*Email:* gsthakur@iiserbpr.ac.in

**We report the synthesis, structure and magnetic properties of two 1:3 ordered quadruple perovskites $Sr_4MRu_3O_{12}$ ($M$ = Li and Na). $Sr_4NaRu_3O_{12}$ crystallizes in the centrosymmetric space group $R\overline{3}$ and $Sr_4LiRu_3O_{12}$ appears to be isostructural to the Na compound based on the PXRD data. In $Sr_4NaRu_3O_{12}$, both Na and Ru are predominantly ordered at the *B* sites (here Na/Li and Ru) and the structure contains only corner-connected $RuO_6$ and $NaO_6$ octahedra. This atomic ordering also leads to a rather large unit cell with $a$ = 11.25 Å and $c$ = 27.6 Å compared to the basic 12R structure ($a$ = 5.5 Å and $c$ ~ 27 Å). Magnetic measurements reveal that $Sr_4NaRu_3O_{12}$ undergoes a magnetic transition to an antiferromagnetic state below $T_N$ ~ 265 K which is confirmed by DSC and neutron diffraction. The Ru moments show a collinear antiferromagnetic spin alignment along the hexagonal $c$ axis with a propagation vector $k$ = (0, 0, 1.5). Interestingly, those Ru moments lying on the three-fold roto-inversion do not significantly contribute to the magnetic order, since they are located between antiferromagnetically coupled Ru atoms and are therefore probably highly frustrated. Band structure calculations on**

**$Sr_4NaRu_3O_{12}$ complement the observed magnetic ground state and a semiconducting behavior in the compound. $Sr_4LiRu_3O_{12}$ shows a magnetic anomaly below 110 K, possibly associated with competing ferromagnetic and antiferromagnetic interactions.**

## Introduction

4*d*- and 5*d*-transition-metal oxides continue to attract considerable attention because their electronic and magnetic ground states arise from a delicate interplay of bandwidth, electron correlation, crystal-field splitting, and spin-orbit coupling.[1] In ruthenates, this competition gives rise to a remarkable diversity of phenomena, including itinerant ferromagnetism in $SrRuO_3$,[2,3] spin-triplet superconductivity in $Sr_2RuO_4$,[4] meta-magnetism in $Sr_3Ru_2O_7$,[5] and complex magnetic behavior in $Sr_4Ru_3O_{10}$.[6] At the same time, ruthenium oxides are also relevant from a chemical perspective because of their catalytic and electrocatalytic properties.[7–9] This combination of fundamental and functional interest makes ruthenates a particularly attractive platform for discovering new structure-property relationships.

Among oxide structure families, perovskites have highly flexible structure because they accommodate a wide range of cations and tolerate extensive structural variation while preserving robust metal-oxygen frameworks.[10–16] In particular, the connectivity of $BO_6$ octahedra strongly influences the electronic and magnetic properties. For 4*d* and 5*d* oxides, extended metal-oxygen-metal interactions can promote strong exchange coupling and, in favorable cases, magnetic ordering temperatures that are unusually high for compounds containing only heavier transition metals.[17–23] Although such behavior is more commonly associated with 3*d* systems, several ruthenates, including $ARu_2O_6$ ($A$ = Sr, Ba) and $AgRuO_3$, have demonstrated that Ru-based oxides can also support antiferromagnetic order at or above room temperature.[24–28] These observations motivate the search for new $Ru^V$ oxides with distinct structural motifs and exchange pathways.

A particularly interesting but only sparsely explored subgroup is formed by 1:3 ordered quadruple perovskites of composition $A_4B'B_3O_{12}$. In these compounds, cation ordering can generate complex superstructures and unusual connectivity of the $BO_6$ octahedra, thereby opening access to magnetic states not realized in simpler perovskites. Known reference systems such as $Ba_4NaRu_3O_{12}$, $Ba_4NaSb_3O_{12}$, and $Sr_4NaSb_3O_{12}$ illustrate that closely related compositions can adopt rather different structure types, ranging from hexagonal variants with face-sharing octahedra to cubic or pseudocubic arrangements.[29–31] Earlier work on $Sr_4(M_{1/4}Ru_{3/4})O_{12}$ ($M$ = Li, Na) described $B$-site-disordered orthorhombic phases obtained under flowing oxygen and characterized them as paramagnetic semiconductors.[32] In contrast, an atomically ordered Sr-based $Ru^V$ quadruple perovskite with well-defined magnetic order has remained elusive.

Here we report the synthesis, crystal structure, and physical properties of $Sr_4NaRu_3O_{12}$ and compare them with those of the Li analogue $Sr_4LiRu_3O_{12}$. $Sr_4NaRu_3O_{12}$ is obtained under comparatively simple solid-state conditions in air and adopts a rare ordered 12$R$ quadruple-perovskite structure with predominant Na/Ru ordering on the $B$ sites and exclusively corner-sharing octahedra. This distinguishes it from the known Ba analogue and from the previously reported disordered Sr phases.[30,32] Most importantly, $Sr_4NaRu_3O_{12}$ exhibits bulk long-range antiferromagnetic order below about 265 K, placing it among the few ruthenates with magnetic ordering close to room temperature. The magnetic ground state is established by susceptibility, calorimetry, and neutron diffraction, while transport and electronic-structure calculations indicate semiconducting behavior with a narrow band gap. For $Sr_4LiRu_3O_{12}$, the diffraction and spectroscopic data are consistent with a closely related $Ru^V$ oxide, and magnetization measurements reveal an anomaly near 110 K that points to a substantially different low-temperature magnetic state. Together, these results show that quaternary pentavalent ruthenates of the $Sr_4MRu_3O_{12}$ type constitute an accessible family of magnetic oxides and an appealing starting point for the discovery of further antiferromagnetic semiconductors relevant to modern oxide chemistry and, potentially, antiferromagnetic spintronics.[33–38]

## Experimental

**Synthesis:** Powder samples of $Sr_4MRu_3O_{12}$ ($M$ = Li and Na) were synthesized by direct solid-state reaction of $SrCO_3$, $M_2CO_3$ and $RuO_2$ in air. Stoichiometric amounts of the starting materials (a total of 0.25 g) were thoroughly ground using a ball mill for 15 minutes and pelletized into an 8 mm disk. The pellets were placed in an alumina crucible and heated up to 1173 K in air for 12 to 48 h at a rate of 100 K/h. The resulting products were porous lump of dark grey and shiny appearance. The products were consolidated into a pellet and heat-treated at 1173 K for 12 h in air to obtain hard-dense pellets. A single-phase sample can be obtained using this method if the total quantity of the product is kept $\leq$ 0.3 g, larger batches tend to include $SrRuO_3$ impurity in amounts detectable by PXRD. Small single crystals of $Sr_4NaRu_3O_{12}$ suitable only for diffraction studies were grown by heating and slow cooling non-stoichiometric amounts of $SrCO_3$, $Na_2CO_3$ and $RuO_2$ in a ratio of 4:10:5. Around 0.5 g of raw materials were pelletized and heated at 900° C for 15 h and then slowly cooled to 800° C in 75 hours. After this the furnace was allowed to cool naturally to room temperature. Crystals were separated by dissolving the flux in distilled water aided with sonication. Tiny hexagonal or block shaped single crystals were picked manually from rest of the powder under optical microscope for single crystal and EDX studies. The crystal growth attempt for $Sr_4LiRu_3O_{12}$ only resulted in very tiny crystals unsuitable for single-crystal diffraction experiments.

**Compositional analysis:** Elemental analysis of consolidated polycrystalline as well as single crystalline sample was carried out by SEM-EDX technique using a SU8020 (Hitachi) equipped with a triple detector system for secondary and low-energy backscattered electrons applying an acceleration voltage of 20 keV for EDX. At least 20 different crystals were screened to ascertain the elemental (metal only) composition. The ratio of Sr:Ru:Na was found to be 4:3.2:0.9. The EDX mapping indicated a uniform distribution of all the elements throughout the crystals (Figure S1 in SI). Due to the limitation of the detection limit of the EDX technique Li could not be detected; however, the Sr/Ru ratio was very close to 1.33. Chemical analysis on powder sample using ICP-OES also confirmed the elemental composition to be very close to $Sr_4(Na/Li)Ru_3O_{12}$.

**Single Crystal X-ray diffraction:** Diffraction data were collected at ambient temperature with a four-circle diffractometer Kappa Apex II (Bruker-AXS, Karlsruhe, Germany) equipped with a CCD-detector using graphite-monochromated Mo-Kα radiation ($\lambda$ = 0.71073 Å). The raw data were corrected for background, Lorentz and polarization factors,[39] and multi-scan absorption correction was applied.[40] The structure was solved using *SUPERFLIP*[41] and the structure model was refined against $F^2$ with *JAN2020*.[42] The graphical representations of the structure were developed with *Diamond*.[43] A summary of the structure data for $Sr_4NaRu_3O_{12}$ are given in Table 1; the information, as well as the atomic coordinates, selected interatomic distances and anisotropic displacement parameters are listed in Tables S1-S4, respectively, in the SI file.

**Table 1**: Results of the crystal structure refinements of $Sr_4NaRu_3O_{12}$.

| | |
|---|---|
| Temperature | 298(2) K |
| Refined formula | $Sr_4Na_{0.84}Ru_{3.06}O_{12}$ |
| Formula weight, g mol$^{-1}$ | 868.8 |
| Space group (no.) | $R\bar{3}$, (148) |
| Unit cell dimensions, Å | $a$ = 11.2332(5); [$a$ = 11.232(1)]*<br>$c$ = 27.532(1); [$c$ = 27.727(1)]* |
| Volume, Å$^3$ | 3008.7(2) |
| Density calculated, g cm$^{-3}$ | 5.771 |
| Data / restraints / parameter | 3142 / 0 /79 |
| Goodness-of-fit on $F^2$ | 1.6 |
| Final $R$ indices [$I > 3\sigma(I)$] | $R_1$ = 0.0664, w$R_2$ = 0.1635 |
| $R$ indices (all data) | $R_1$ = 0.1548, w$R_2$ = 0.1895 |
| Largest diff. peak and hole (e·Å$^{-3}$) | 5.02/–6.24 |

*Values obtained from PXRD data.

**Powder X-ray and neutron diffraction:** Room temperature PXRD data was collected on Malvern Panalytical Empyrean 3 diffractometer equipped with a copper K$\alpha$ source ($\lambda$ = 1.5406 Å). The diffraction data was collected in the 2$\theta$ range of 5° to 90°, with a step size of 0.013 and 3D pixel detector in Bragg-Brentano setup. Neutron powder patterns were collected on the instrument D20 at the Institute Laue-Langevin (ILL)[44] using a Ge(113) monochromator selecting the neutron wavelength $\lambda$ = 2.41 Å. A large number of powder patterns in the scattering range from 2$\theta$ = 4 to 150° were collected in the temperature from 3 to 300 K. Rietveld refinements of the powder diffraction data were carried out with the program *FullProf*.[45] For the crystal structure refinements the nuclear scattering lengths $b$(O) = 5.804 fm, $b$(Na) = 3.64 fm, $b$(Sr) = 7.02 fm, and $b$(Ru) = 7.03 fm were used.[46] The magnetic form factors of the Ru atoms were taken from.[46,47] It should be noted that the magnetic form factor of the ion $Ru^{V}$ is available.[48]

**Electrical resistivity:** DC resistivity of a sintered pellet of $Sr_4NaRu_3O_{12}$ was measured on an approximate bar shaped sample using four-probe method in the temperature range 75 to 300 K. Electrical contacts were made with conducting silver paste using 25-micron thick Pt wire. The sample was cooled in a cryogen free measurement system (CFMS, Cryogenic Ltd.), Keithley 2400 source meter and a Keithley 2182A nanovoltmeter were used to supply the current and record the voltage, respectively.

**Magnetic susceptibility:** Magnetization measurements on several polycrystalline samples of $Sr_4NaRu_3O_{12}$ were recorded on a Vibrating Sample Magnetometer (VSM) using a cryogen free measurement system (CFMS, Cryogenic Ltd.). Magnetic susceptibility was measured in the temperature range 2 to 300 K at applied magnetic field between 0.1 to 9 T. Isothermal magnetization at 2 K was measured in an applied magnetic field up to 9 T. High temperature magnetic susceptibility in the temperature range 300 to 750 K was measured using a VSM equipped with an oven. For the $Sr_4LiRu_3O_{12}$ sample, magnetic measurements were performed using a MPMS-3 (Quantum Design,) in the temperature range 1.8 to 300 K and in different magnetic fields. High temperature magnetic susceptibility in the range 300 to 750 K was measured using the VSM attachment in PPMS equipped with an oven.

**Heat Capacity:** Temperature-dependent heat capacity [$C_p(T)$] was measured down to 2 K on a sintered pellet in a PPMS (Quantum Design) using the standard thermal relaxation technique of the HC option.

**DFT Calculations**: We have used DFT calculations with pseudopotentials based on the projected augmented wave (PAW)[49] method as implemented in the Quantum Espresso (QE) package.[50] The exchange-correlation potentials are described with Generalized Gradient Approximations (GGA)[51] with Perdew–Burke–Ernzerhof (PBE) parameterization. The Monkhorst–Pack recipe[52] was used for *k*-points sampling with a Gamma centered grid 4 × 4 × 4, and plane wave energy cut-off was set to 75 Ry along with a total energy convergence threshold of $8.0 \times 10^{-5}$ Ry. The crystallographic details used for the calculation are obtained from the solved structure of SCXRD data. The magnetic ground state and electronic band structure were calculated in a collinear fashion with an effective Hubbard U energy value of 2.0 eV for the Ru 4*d* orbitals following the Dudarev et al. approach.[53]

## Results and Discussion

Air stable dark grey powder samples of $Sr_4MRu_3O_{12}$ were prepared using a solid-state method. The compound $Sr_4NaRu_3O_{12}$ is robust against thermal decomposition in air until 1000 K (Figure S2 in SI). The samples appeared phase pure as adjudged by PXRD data; however, a trace of $SrRuO_3$ always persisted as detected in the magnetic studies. Second annealing at 1173 K did not improve the final products. It is worth noting the single crystals of some Ba-containing ruthenates can be easily obtained by using hydroxide flux.[54–56] This method however, failed to yield single crystals of Sr compound due to the ease of formation of competing phases $Sr_3NaRuO_6$ and $SrRuO_3$.

**Structure:** The PXRD patterns of $Sr_4NaRu_3O_{12}$ have a remarkable resemblance to $Na_{1.5}Ag_{1.5}MoO_3F_3$,[57] which is an ordered $LiNbO_3$ type structure ($R\bar{3}$) with lattice parameters $a$ = 5.7 Å and $c$ = 28.5 Å (Figure S3 in SI). An initial indexing in similar cell with $a$ = 5.6 Å and $c$ = 27.7 Å, gave good results but some reflections with weak intensity remained unindexed. However, a unit cell with doubled $a$ parameter could account for all the reflections observed in the PXRD pattern (Figure 1). Le Bail fit to the powder patterns for indexed in a rhombohedral cell with lattice parameters $a$ = 11.142(1), $c$ = 27.176(3) (for $Sr_4LiRu_3O_{12}$) and $a$ = 11.232(1) Å, $c$ = 27.727(1) Å (for $Sr_4NaRu_3O_{12}$) are shown in Figure 1. Indexing and crystal structure of $Sr_4NaRu_3O_{12}$ were confirmed by single crystal data collected at 300 and 100 K and the structure was solved using 300 K crystal data. The structure belongs to a rare 12R quadruple-perovskite with *B*-site ordering between Na and Ru in 1:3 ratio. A small degree of anti-site disorder between the Na and Ru sites was also observed during the structure refinement against single crystal data, see below. The final composition from single crystal data was refined to $Sr_4Na_{0.84}Ru_{3.06}O_{12}$. This composition is close to nominal and to what determined from EDX and ICP analysis.

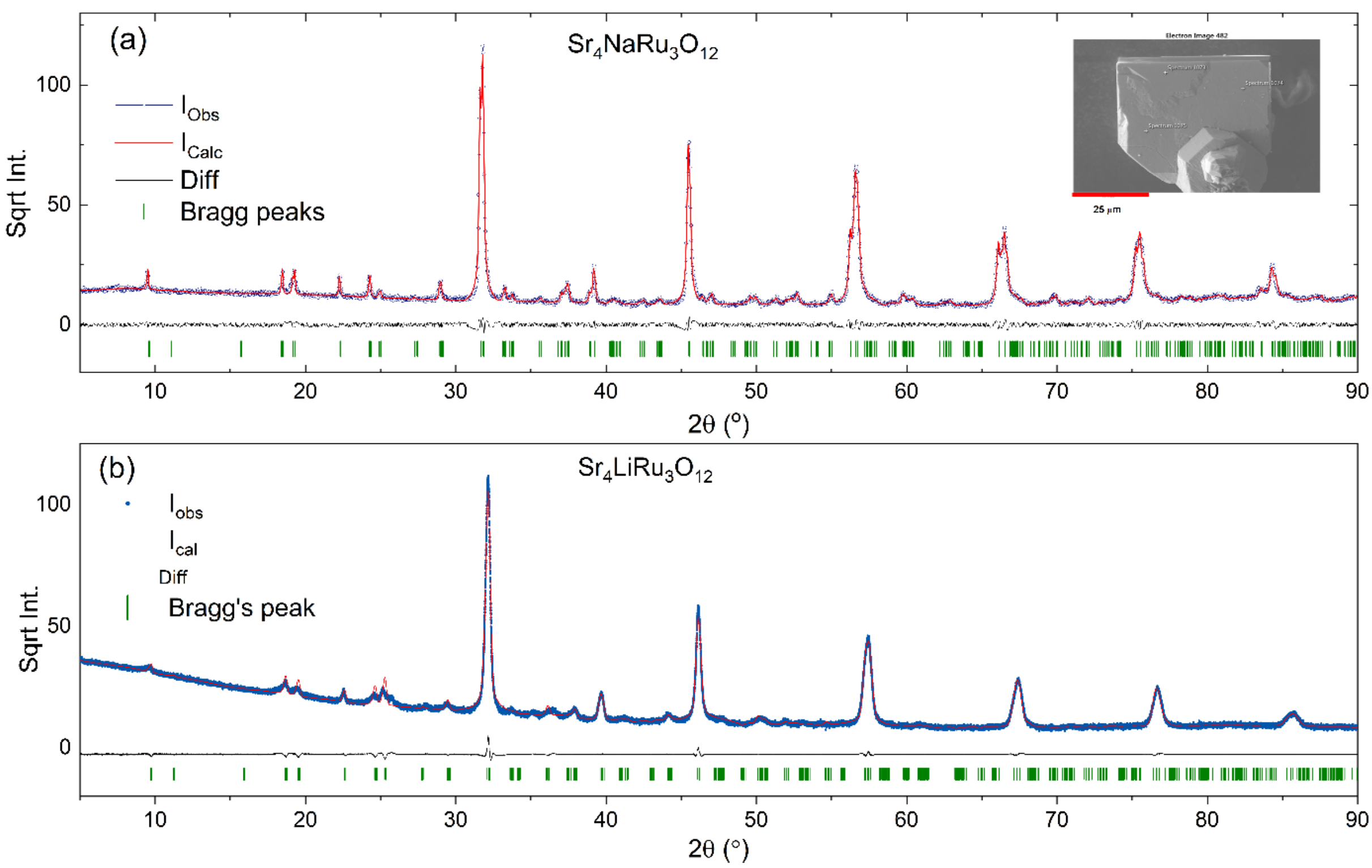

**Figure 1.** Le Bail fit of room temperature PXRD data of $Sr_4NaRu_3O_{12}$ (a) and $Sr_4LiRu_3O_{12}$ (b). Blue and red curves are observed and calculated intensities and black line is their difference. Green vertical bars indicate allowed reflections. Inset of (a) shows the SEM image of a typical $Sr_4NaRu_3O_{12}$ crystal.

The crystal structure of $Sr_4NaRu_3O_{12}$ is displayed in Figure 2 in different crystallographic orientations. The crystal structure is a rhombohedral stack of 12 layers of metal-octahedra that can be considered as a section of cubic perovskite cut perpendicular to one of its 3-fold axes. It is largely made up of three layers of corner shared $RuO_6$-octahedra which is interspersed by a layer of $NaO_6$ octahedra. However, this ideal triple layer arrangement is interrupted as Na occupy some sites in the Ru triple layer and *vice-versa*. The ruthenium atoms are distributed over six distinct crystallographic sites (3*a*, 18*f*, 9*d*, 6*c*, 3*b* and 9e), out of which four (3*a*, 18*f*, 9*d*, 6*c*) are exclusively occupied by Ru atoms whereas the remaining two (3*b* and 9*e*) are mostly occupied by Na atoms (> 75 %). Ruthenium atoms at the 3*a*, 18*f*, 9*d*, and 6*c* sites have an octahedral coordination with Ru–O distances ranging between 1.85 and 2.1 Å (Table S3). The two positions mainly occupied by Na atoms (labelled Na1 and Na2 in the following) have longer *M*-O distances in the range 2.1-2.2 Å. The bond valence sum (BVS) for Na1 and Na2 was calculated to 2.2 and 2.0, respectively. The bond valence sums and the interatomic distances both support a mixed site occupation scenario. The observed distances Ru–O *in principle* agree with literature values for Ru(V).[27,30,58–60]

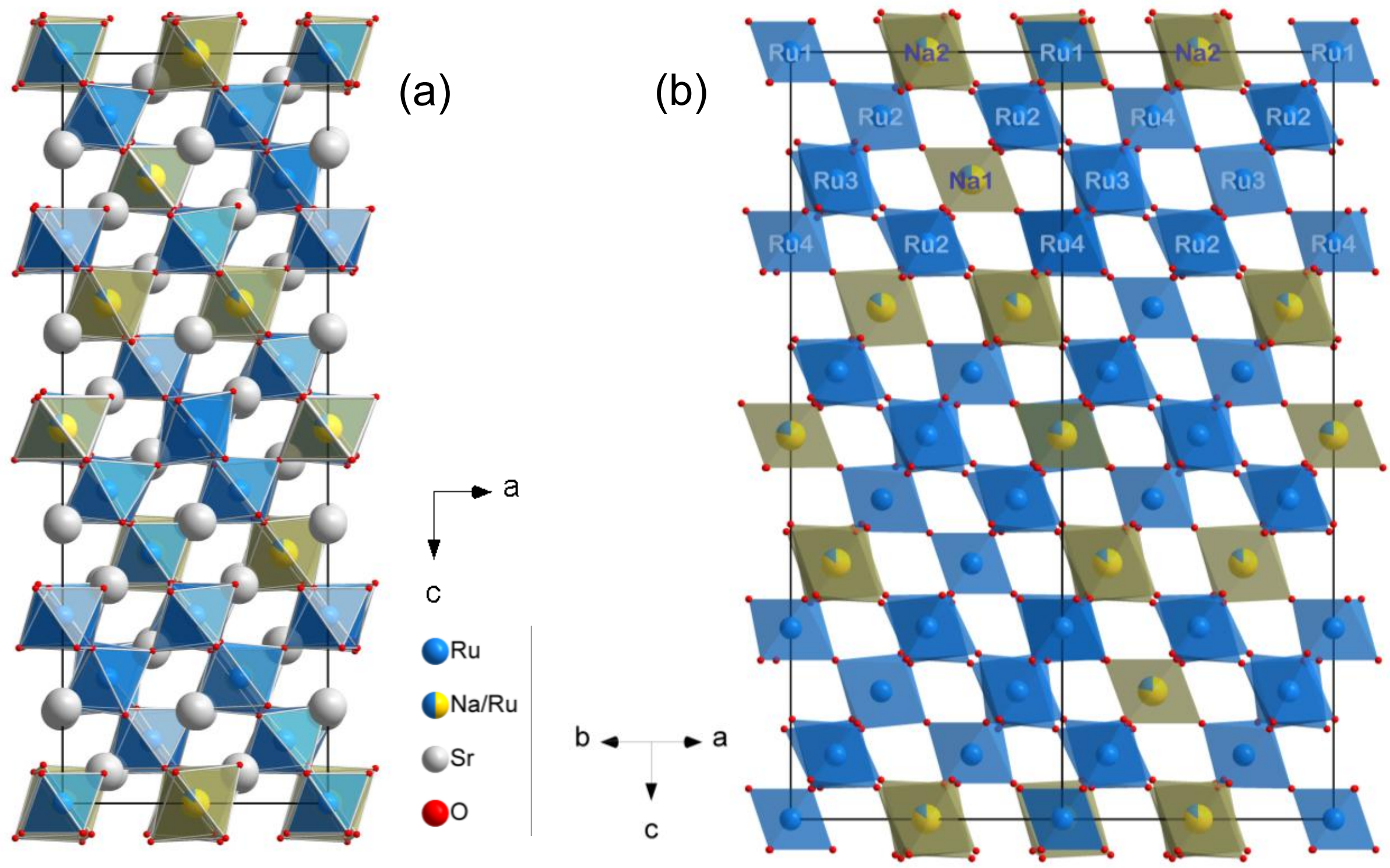

**Figure 2**. Crystal structure of $Sr_4NaRu_3O_{12}$ viewed along different orientations (*a* and *b*).

The Ru(1)$O_6$ octahedra is fairly regular with distances of approximately 1.87 Å. The Ru(2)$O_6$ and Ru(4)$O_6$ have three short and three long contacts which leads to a slight off-centering of the Ru atom towards one of the triangular faces of the octahedra. The bond valence sums for Ru2 and Ru3 are close to 5 whereas for Ru1 and Ru4 it is 5.5 and 5.6, respectively.

Interestingly, all the *B*-site metal octahedral are only corner shared to each other and no face sharing is observed. It is worth noting that the Ba-analogue $Ba_4NaRu_3O_{12}$ is reported and possesses 8H perovskite structure with (*chhh*)$_2$-layer sequence and consisting of a face-shared $Ru_2O_9$ dimers.[30] However, $Sr_4NaRu_3O_{12}$ adopts a different 12 layered structure with only corner sharing octahedral connectivity *akin* to a conventional perovskite. In Ba-containing perovskites, proportions of face-sharing octahedra are usually governed by the tolerance factor($\tau$).[30,61,62] $\tau$ value higher than unity leads to an increasing proportion of hexagonal layers (face sharing) in the structure. However, $\tau \sim 1$ should lead to structures with high proportions of cubic layers (corner sharing), which is the case for the present compound with $\tau \sim 0.98$. It may be expected that the structure of $(Sr/Ba)_4NaRu_3O_{12}$ would gradually change from only corner sharing to 8H perovskite with structurally interesting intermediates. To test this, we synthesized the compounds of composition $(Sr_{1-x}Ba_x)_4NaRu_3O_{12}$ ($x$ = 0 to 1). We observed that the solubility limit of Ba in Sr compound (or vice-versa) is approximately 30 %. With $x$ = 0.333 to 0.666, phase separation occurs and a mixture of the two phases (rhombohedral and hexagonal) are observed (Figure S4 in the SI). These structural features also make $Sr_4NaRu_3O_{12}$ structurally distinct, as all other known 1:3 ordered quadruple perovskites with either *A* or *B* site order crystallize in either hexagonal system ($A_4B'B_3O_{12}$) or (pseudo-) cubic system ($A'A_3B_4O_{12}$): where the hexagonal phase contains face-sharing *B* metal octahedra and the cubic phase displays square-planes of *A'* cations, respectively.

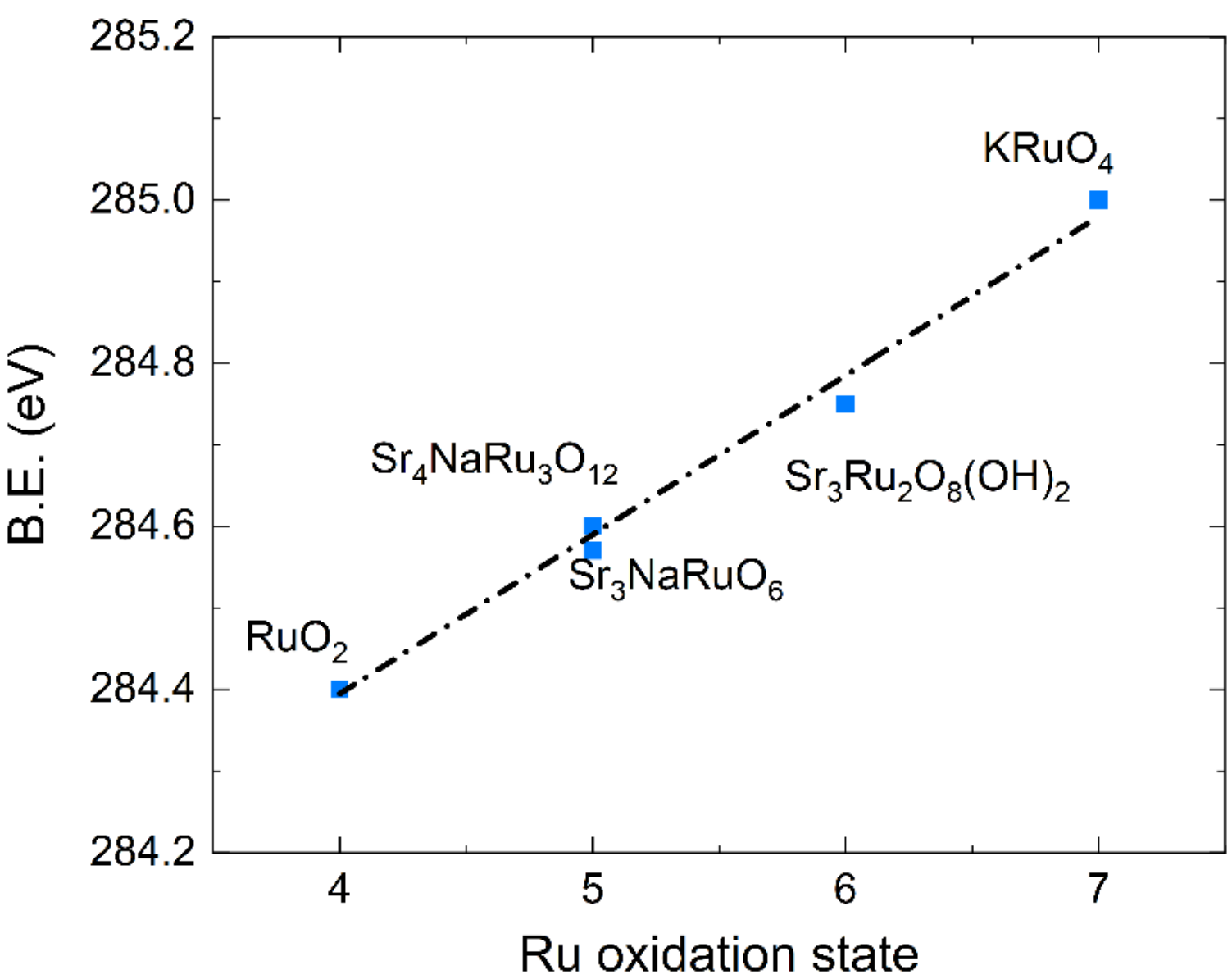

**Figure 3.** XPS data of $Sr_4NaRu_3O_{12}$. The binding energy (B.E.) value for Ru(VI) is taken from $Sr_3Ru_2O_8(OH)_2$, which has been recently confirmed experimentally.[63,64] XPS spectra of $RuO_2$ and $KRuO_4$ were recorded on commercially available samples.

From the nominal composition, an oxidation state close to +5 is expected for ruthenium. To confirm this, we performed XPS studies, the detailed results of which are presented in Figure S5. Given the difficulty in determining oxidation states from absolute binding energies in XPS data due to the overlap of C-1*s* peaks at the same position as Ru-3*d* peaks (Figure S5b), we collected the data for several compounds with Ru in various confirmed oxidation states. A plot of binding energy (B.E.) versus oxidation states follows linearity, and the binding energy for Ru 3*d* in $Sr_4NaRu_3O_{12}$ falls at the expected place on the curve, in turn, confirming the +5 state of ruthenium (Figure 3). The pentavalent nature can be further confirmed from Ru-3*p* spectra, which are less sensitive than Ru-3*d* but can be deconvoluted and fitted (presented in Figure S5c). Similarly, XPS spectra of $Sr_4LiRu_3O_{12}$ also show a similar set of peaks corresponding to Ru-3*p* and Sr-3*d*, confirming a similar oxidation state to the Na analogue. It may be noted that due to its low electron count, Li is not very sensitive to XPS; hence, we only observed a broad hump corresponding to the Li-binding energy (Figure S5(h)), which is insufficient to deconvolute and fit; however, it suggests the presence of Li in our compound.

The resistivity data for $Sr_4NaRu_3O_{12}$ collected in a temperature range of 80 - 300 K is presented in Figure 4. The electrical resistivity increases non-linearly on cooling and the derivative shows a dip at 260 K. The high-temperature part of $ln\rho$ versus $T^{-1}$ can be described only over a limited temperature interval and yields a small apparent activation energy of about 50 meV. This value should not be equated directly with the intrinsic band gap, because the full temperature dependence is better described by 3D variable-range hopping, likely reflecting disorder- or defect-assisted transport. The transport data therefore support semiconducting behavior, whereas the PBE calculation gives a small electronic band gap of about 250 meV (discussed later).

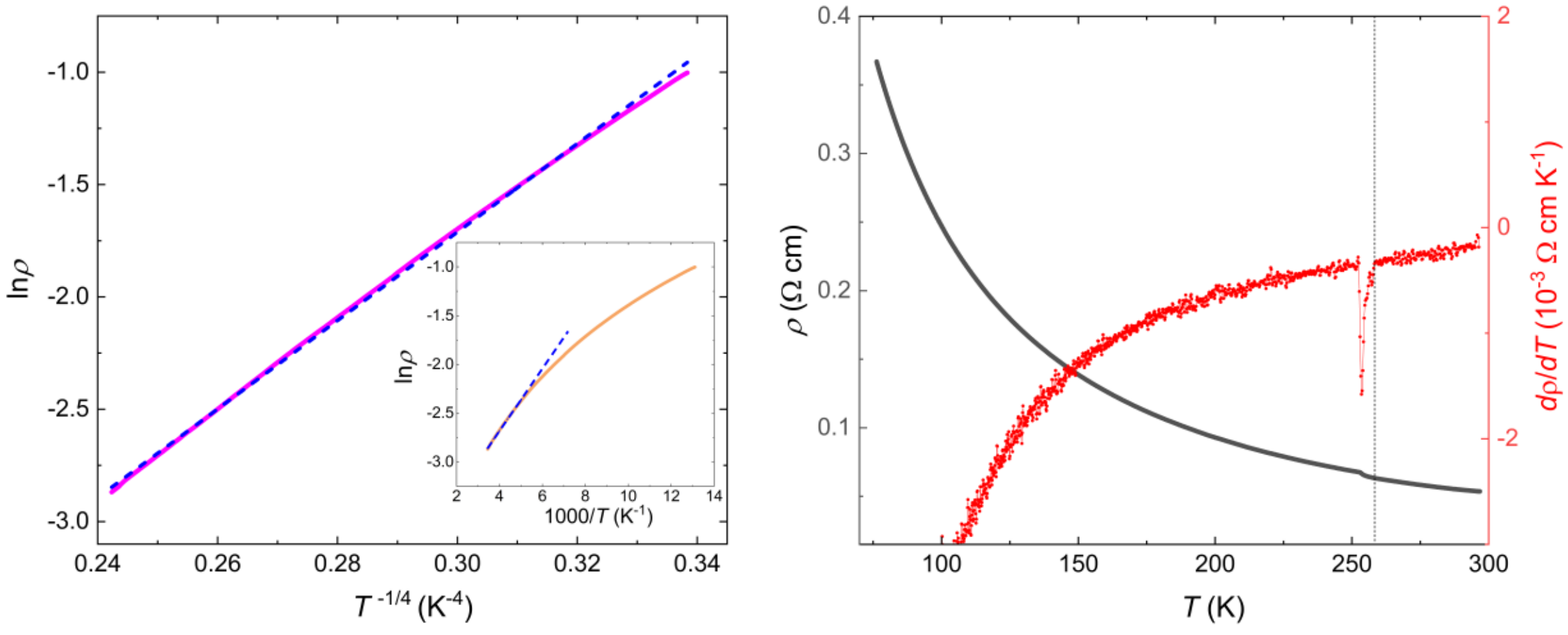

**Figure 4.** (a) Logarithm of resistivity plotted against $T^{-1/4}$ for $Sr_4NaRu_3O_{12}$; inset displays the ln$\rho$ vs $T^{-1}$ curve (b) $\rho$ vs $T$ curve and its first derivative exhibiting a sharp dip at $T$ ~ 260 K.

**Magnetic properties of $Sr_4NaRu_3O_{12}$.**The temperature dependence of molar magnetic susceptibility and isothermal magnetization loop is displayed in figure 5. Magnetic susceptibility measured in the temperature-range from 2 to 300 K at all applied fields (0.1 to 7 T) showed a small but discernible anomaly at 270 K, which is indicative of magnetic ordering (figure 5a). Néel temperature $T_N$ = 265 K was deduced from the first derivative of $\chi$(T) data. The data were corrected for diamagnetic contribution of the individual ions.[65] At higher temperatures (300 to 700 K), it shows no more transition and behaves as a paramagnet.

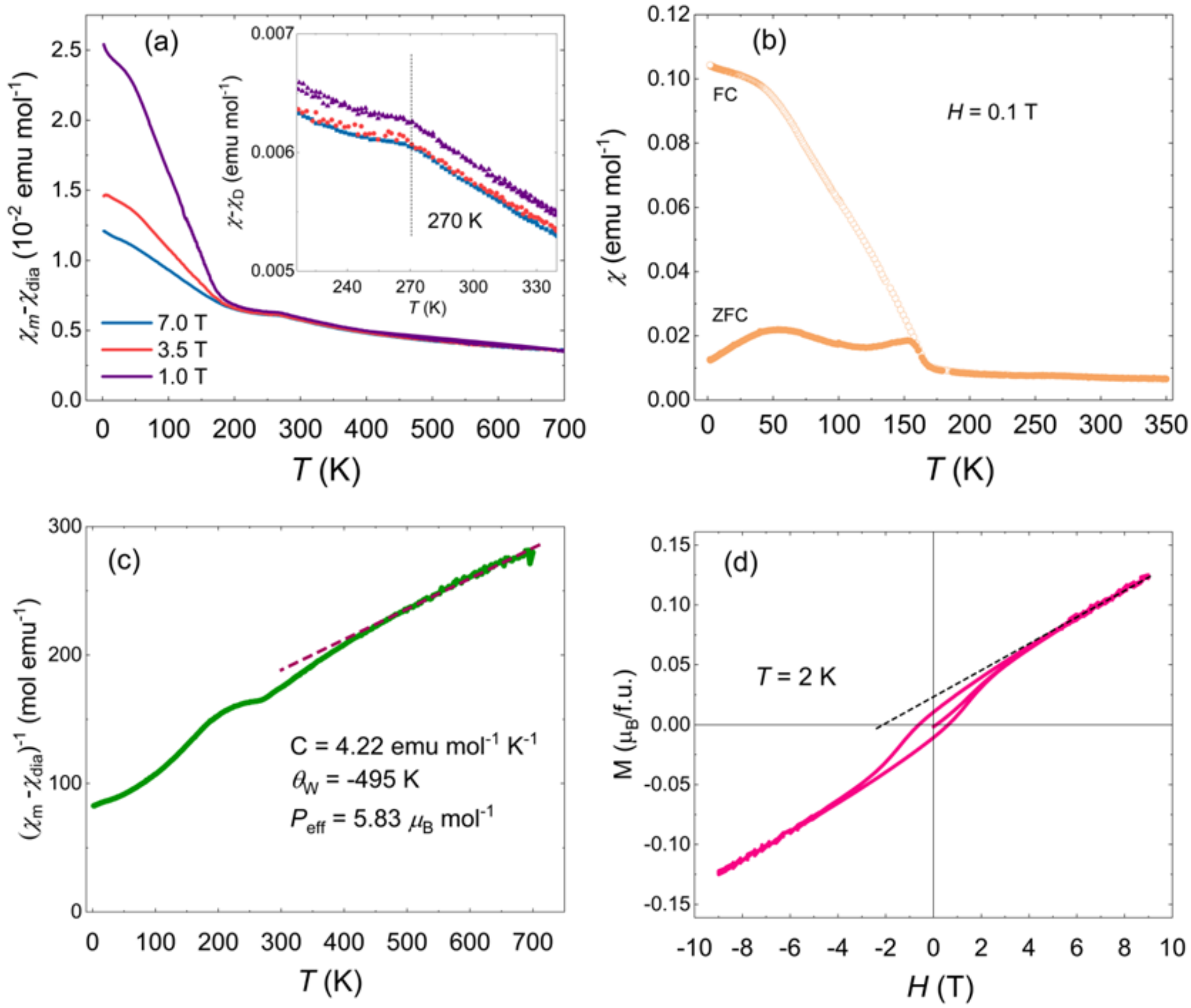

**Figure 5:** Magnetic properties of $Sr_4NaRu_3O_{12}$: (a) magnetic susceptibility in the *T*-range 2 - 700 K at different applied field, inset showing the magnified region near the magnetic transition (b) magnetic susceptibility in ZFC and FC protocols at an applied field of 0.1 T (c) straight line fit to the inverse susceptibility (d) magnetic hysteresis loop collected at 2 K.

A small increase in susceptibility below 170 K, only prominent in the low field data, is also observed which can be most possibly be attributed to the ferromagnetic transition of $SrRuO_3$ ($T_C$ ~ 165 K) (figure 5b). Overall linear behavior of magnetization with applied magnetic field and an extremely small moment at low temperature also supports antiferromagnetic nature of the sample. Small coercivity and remnant moment is observed in the otherwise linear hysteresis curve confirming the presence of trace ferromagnetic impurity $SrRuO_3$ (figure 5d). An upper estimate of ~1 % of this impurity was made based on the experimentally reported saturated moment of $SrRuO_3$ (~1.6 $\mu_B$).[66] Such small amount of secondary phase is impossible to detect in PXRD.

To extract meaningful information on the type and strength of magnetic interactions in the paramagnetic regime, a linear fit to the high temperature (above 2 × $T_N$) region of inverse susceptibility was performed (figure 5c). A straight-line fit yielded a Weiss constant, $\theta_{cw}$ of −540 K indicating strong antiferromagnetic correlations between the spins concomitant with a high magnetic ordering temperature. The effective paramagnetic moment of 3.4 $\mu_B$ per $Ru^{5+}$ ion is calculated from the fit. This value is arguably smaller than the ideal spin-only value of 3.83 $\mu_B$ expected for a $d^3$ system. Similar smaller than expected spin-only value of moments has been reported for other $Ru^{5+}$ oxides, which was partially attributed to a certain degree of covalency of the Ru–O bonds.[24,28,67,68,69] However, spin-orbit coupling could also be the source for a reduced magnetic moment. For the $t_{2g}$ configuration in an octahedral ligand field the magnetic moment is reduced by the factor ($1 - a\lambda/10Dq$) from the spin only value. A typical value of $\lambda$ = 500 $cm^{-1}$ for $Ru^{5+}$ ion leads to a reduced spin-only magnetic moment of ~3.5 $\mu_B$ which is close to the observed value. The antiferromagnetic transition below 270 K is confirmed in DSC studies were the curve shows a sharp, single prominent peak arising at 262 K (figure S6 in SI). A slight mismatch in the temperature of observed transitions between the magnetic and DSC studies might be due to the faster scan rate typically used in DSC measurements. Further, the antiferromagnetic nature of the magnetic ordering was confirmed by neutron diffraction studies discussed in the later section. It is to be noted that a high ordering temperature in only Ru containing oxides is rare and has been observed previously only in few structurally related oxides $ARu_2O_6$ ($A$ = Sr and Ba) and $AgRuO_3$.[25–28] This is partly because of smaller correlation and a larger band width as compared to 3*d*-metal containing compounds, however, Sarma *et al.* have proposed that oxides with $4d^3/5d^3$ can become candidate materials to show high magnetic ordering temperatures.[70]

**$Sr_4LiRu_3O_{12}$**. The magnetization data on the $Sr_4LiRu_3O_{12}$ sample are presented in figure 6. The magnetic susceptibility in figure 6a reveals a transition ~110 K which is suppressed slightly with increasing field. The magnetic susceptibility measured under ZFC and FC conditions show a clear bifurcation near the transition temperature. Additionally, the *M-H* curve measured at $T$ = 1.8 K (Figure 6c) shows a small hysteresis indicating the presence of a ferromagnetic component in the system. We have fitted inverse susceptibility $1/\chi(T)$ data at $\mu_0 H = 0.5$ T above 400 K using modified Curie-Weiss law:

$$\chi = \chi_0 + \frac{C}{T - \theta_{cw}}$$

Here, $\chi_0$ is the temperature independent contribution in magnetic susceptibility, which includes core diamagnetism and Van-Vleck paramagnetism and $C$ is the Curie constant. The fit yields $\chi_0 \simeq 7.6 \times 10^{-5}$ cm$^3$/mol, $C \simeq 1.76$ cm$^3$-K/mol and $\theta_{cw} \simeq -1034$ K. This large negative $\theta_{cw}$ indicates dominant AFM interactions which is in sharp contrast with the bifurcation in $\chi(T)$ as well as the hysteresis in *M*(*H*) curve. To verify the presence of competing AFM and FM interactions, we plotted $\chi T$ vs *T* (inset of Figure 6b). Usually for a purely FM system $\chi T$ will monotonically increase, while for a purely AFM system, it will decrease monotonically. But in this case, as the temperature goes down, $\chi T$ steadily decreases down to the transition temperature. But near the transition temperature, it shows an upturn followed by a peak and then steadily decreased down to the lowest temperature. This distinct behavior corroborates competing FM-AFM interactions in the system,[71] which might lead to a canted AFM transition.[72] Alternatively, a spin glass transition may also give rise to this kind of behavior.[73,74] The Curie constant, $C$ obtained from the Curie-Weiss-fit corresponds to a paramagnetic moment of $\mu_{\text{eff}} = 3.75\ \mu_B$/Ru$^{\text{V}}$, which is closer to the theoretical spin-only value of $3.87\ \mu_B$ for $S$ = 3/2 with $g$ = 2. In addition, the absence of magnetic anomalies near 160 K indicates the absence of $SrRuO_3$ impurity in the sample. However, $C_p(T)$ data of $Sr_4LiRu_3O_{12}$ shown in figure 6(d) do not show any clear anomaly at the expected magnetic transition temperature. Since the transition is at a relatively high temperature, the corresponding heat capacity anomaly might be suppressed by the phonon contributions. Thus, the susceptibility and hysteresis suggest competing AFM/FM interactions or canted/glassy behavior, but the absence of a heat-capacity anomaly means that a bulk long-range transition is not established.

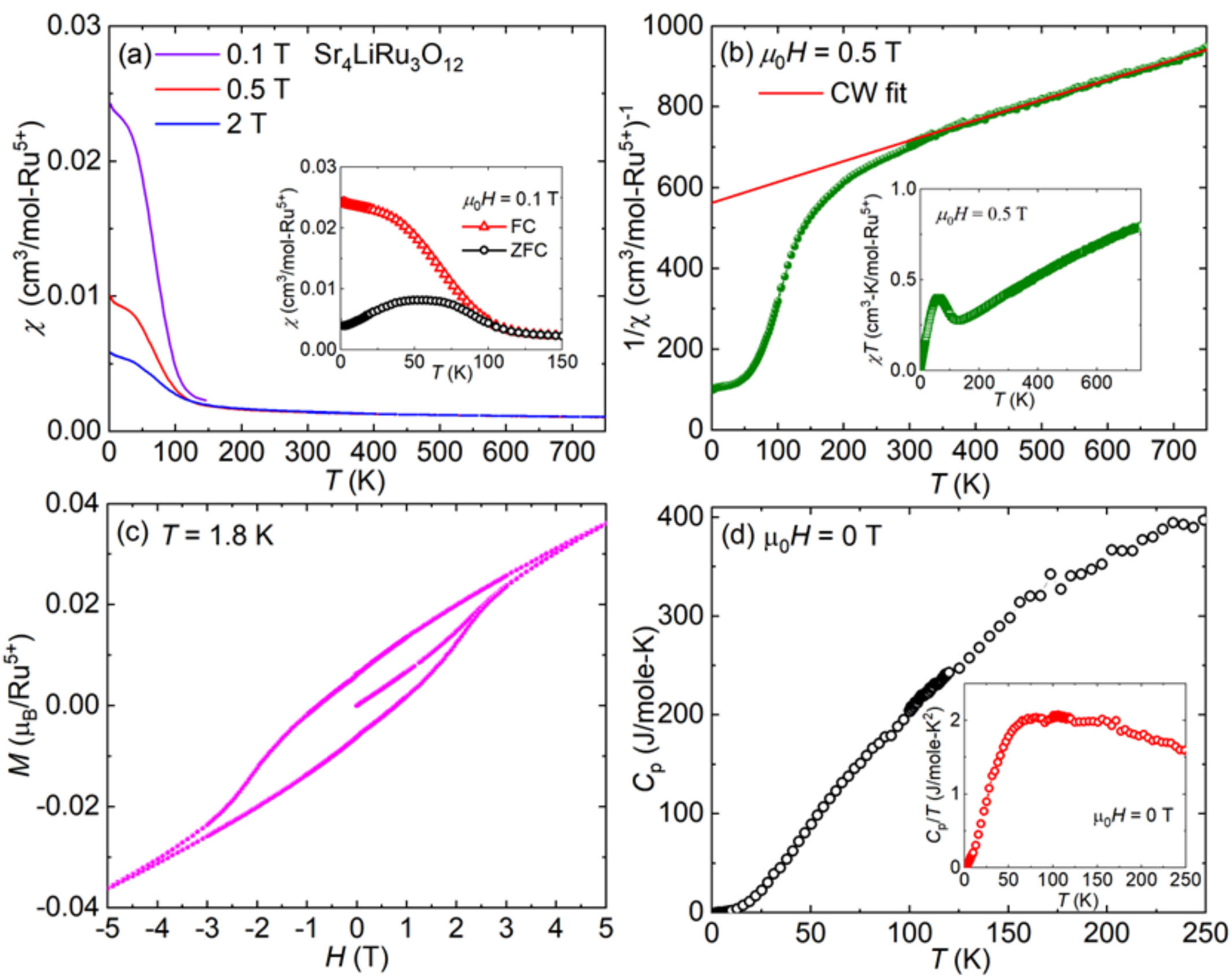

**Figure 6.** Magnetic properties and heat capacity of $Sr_4LiRu_3O_{12}$: (a) magnetic susceptibility in the *T*-range 2 - 750 K at different applied fields, inset showing the magnified ZFC and FC data near the magnetic transition. (b) Straight line fit to the inverse susceptibility at $\mu_0 H = 0.5$ T. Inset shows $\chi T$ vs *T* at the same field. (c) Magnetic hysteresis loop collected at 1.8 K. (d) $C_p$ vs *T* data collected at zero applied magnetic field and inset shows the $C_p/T$ vs *T* plot.

**Heat capacity of $Sr_4NaRu_3O_{12}$:** Molar heat capacity measured as a function of temperature for $Sr_4NaRu_3O_{12}$ is presented in figure 7. Between 2 - 250 K the $C_p(T)$ curve is featureless but a small but noticeable hump is observed at $T$ = 264 K indicates a phase transition. However, it should be noted that the grease (Apiexon-N) used to fix the sample onto the heat capacity puck also contributes greatly in the temperature range 250 - 300 K and thus this upturn may also be just an artefact. However, given the signatures in susceptibility, DSC and neutron diffraction data observed in the similar $T$-range (260 K - 265 K), this may be taken as an evidence of magnetic transition. Nevertheless, this feature is not a typical sharp peak (λ-shape) associated with a phase transition possibly because of small entropy change associated with this particular magnetic phase transition. This is also evident from low temperature diffraction data which indicates that the material does not undergo any structural transition or symmetry lowering. The $C_p(T)$ vs $T^2$ plot was used to estimate the electronic and lattice contribution to the heat capacity. A straight line fit to the low temperature region (< 10 K) of the plot yields $\gamma$ = 15.2 mJ mol$^{-1}$ K$^{-2}$ and $\beta$ = 0.64 mJ mol$^{-1}$ K$^{-4}$. Though the value of $\beta$ is expectedly small, a sizeable value of $\gamma$, in a narrow-gap or hopping semiconductor may arise from atomic site disorder.

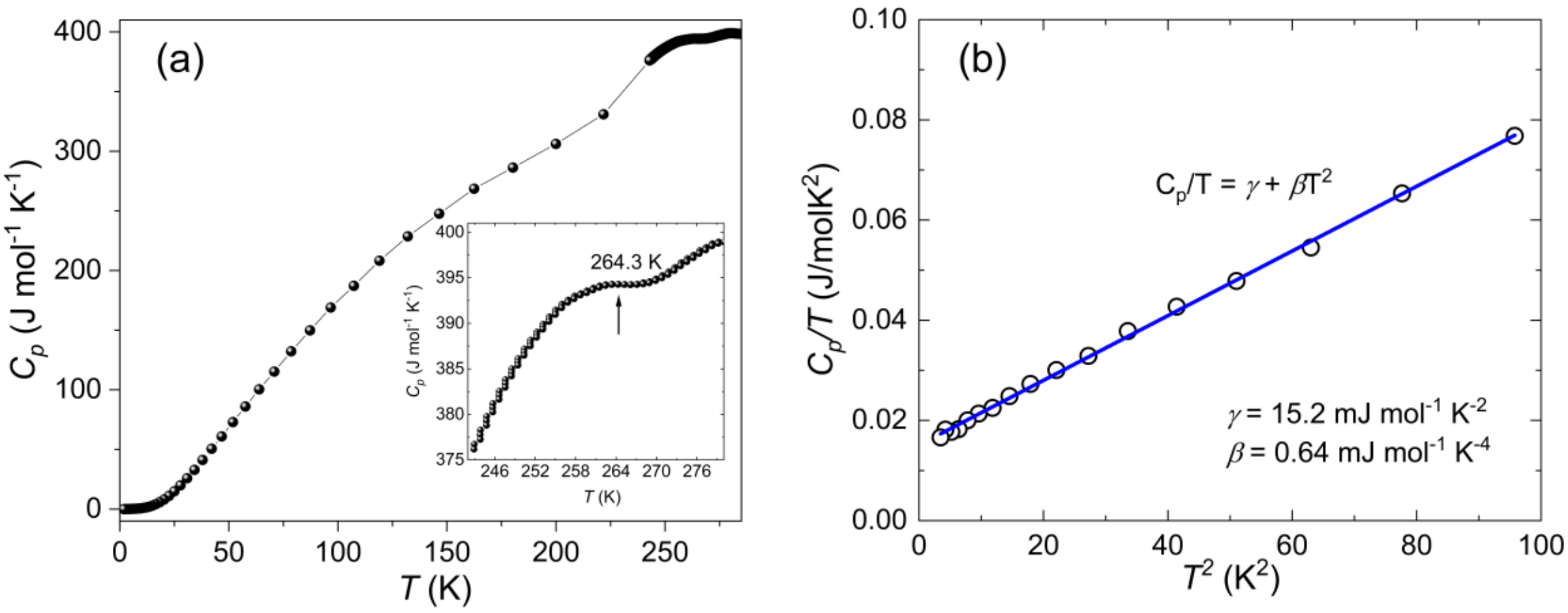

Figure 7 (a) $C_p(T)$ data in the temperature range 2 - 290 K. The inset shows the magnified view near the expected phase transition. (b) $C_p/T$ vs $T^2$ plot and its linear fit in the low temperature region (< 10 K).

**Neutron diffraction of $Sr_4NaRu_3O_{12}$:** In our neutron diffraction study we have investigated the long-range magnetic order of $Sr_4NaRu_3O_{12}$ from data sets collected between 3 and 300 K well below and above the antiferromagnetic ordering temperature $T_N$ = 268 K. Only two magnetic reflections (at 28.9 ° and 34.0 °) could be clearly observed in the neutron diffraction pattern collected at 2 K. Due to the presence of some impurity peaks with similar intensities, pure magnetic contribution becomes more visible in the difference plot 3 K - 300 K shown in Figure 8. The evolution of the intensity of the peak at $2\theta = 28.9^{o}$ with temperature confirms the magnetic origin of this peak and strongly suggests a long-range magnetic ordering in $Sr_4NaRu_3O_{12}$. A closer inspection showed that the reflection observed at 34.0° is not compatible with the space group $R\overline{3}$ and the given lattice parameters, indicating a magnetic propagation vector $\boldsymbol{k} \neq 0$. Both reflections can, however, be indexed using the vector $\boldsymbol{k}$ = (0, 0, 1.5), where the satellites at 28.9 ° and 34.0 ° are indexed as $(2\ 0\ ½)_M$ and $(2\ 0\ 3½)_M$. This propagation vector leads to a doubling of the crystallographic unit cell along the *c* axis, where a spin inversion occurs at all magnetic positions in the next neighboring unit cell along the *c*-axis as shown in figure 8. The magnetic moments in the sublattices of Ru2, and Ru4 (site 18*f* and 6*c*, respectively), are ferromagnetically aligned, while the spins of the Ru3 (9*d*) atoms are coupled antiparallel with respect to Ru2 and Ru4 atoms giving an overall antiferromagnetic coupling. Using this model, we obtained a satisfactory fit, where the individual atoms Ru2, Ru3 and Ru4 reach similar moment values of about 2 $\mu_B$ although affected with increased standard deviations, which can be explained by the fact, that only a small data set was available for the refinements. Refined moment for Ru1 resulted in a value close to zero. Ru1 atoms do not significantly contribute to the magnetic order, since they are located in between two antiferromagnetically coupled layers of Ru atoms and are therefore, probably highly frustrated. Thus, one finds magnetically ordered triple layers lying perpendicular to the *c* axis forming alternate spin sequences + − + , − + −, … along *c* as shown in Figure 8c.

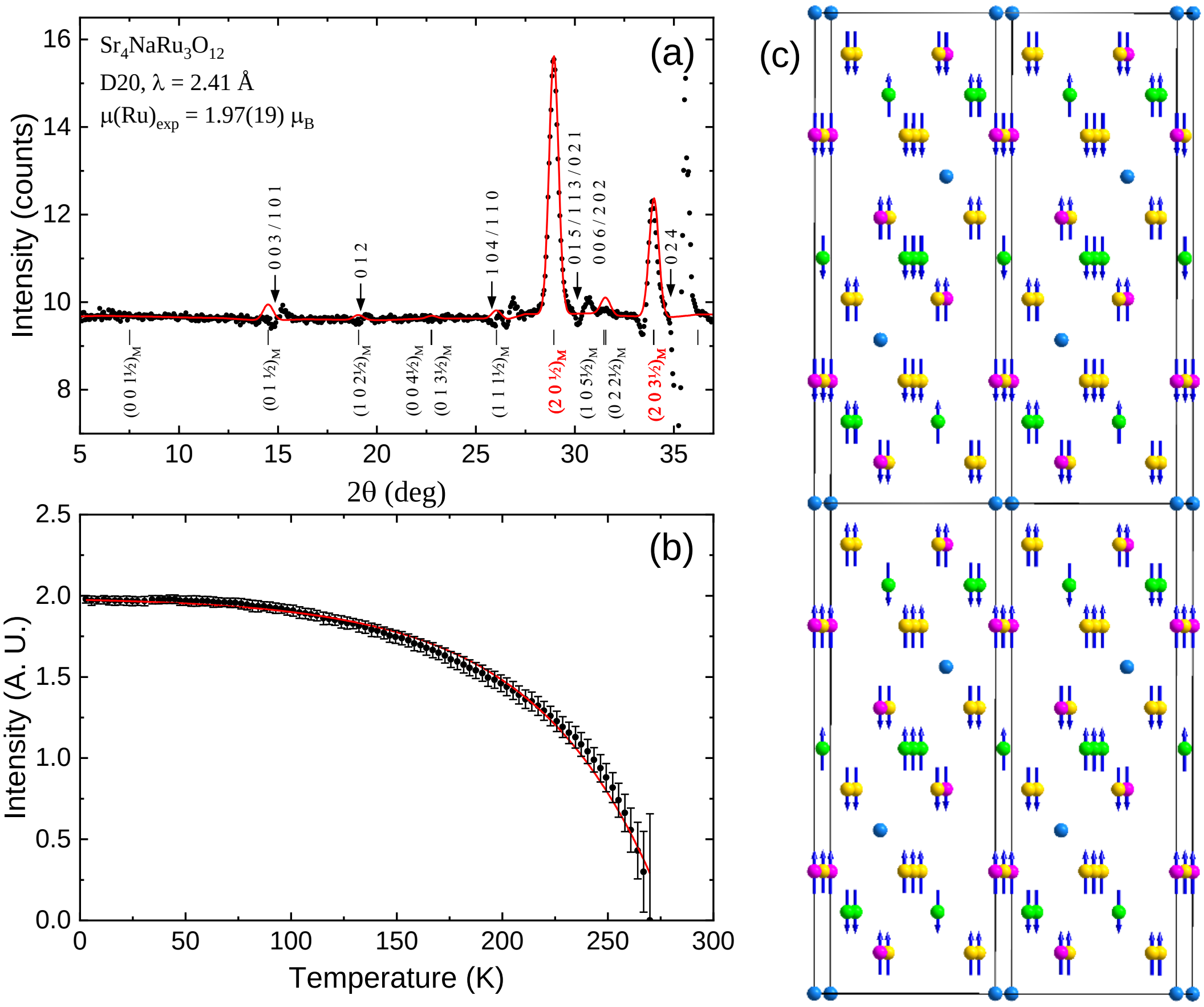

**Figure 8**. Results of the analysis of the powder neutron diffraction data (a) Difference pattern of $Sr_4NaRu_3O_{12}$ obtained from neutron data sets collected at 3 and 300 K. The calculated pattern (red line) is compared with the observed one (black circles). Below the difference patterns the position of magnetic satellites (black bars) are shown. These are generated by the rule $(h'k'l')_M = (hkl)_N \pm \boldsymbol{k}$, where the propagation vector is $\boldsymbol{k} = (0, 0, 1.5)$. The arrows above the pattern mark the positions of the residual intensity of the much stronger nuclear Bragg reflections. Due to the thermal decrease of the lattice parameters down to 2 K, residual intensities of stronger Bragg reflections form zigzag lines which are located close to the positions of the magnetic satellites. (b) Temperature dependence of the intensity magnetic peak at 2θ = 28.9°. (c) Magnetic structure of $Sr_4NaRu_3O_{12}$. The Ru1 atoms (blue), do not contribute to the magnetic order, since they are located between antiferromagnetically coupled Ru atoms. Therefore, one finds magnetically ordered triple layers forming alternate spin sequences + − +, − + − … along the *c* axis. In these layers the Ru2, Ru3 and Ru4 atoms are marked in yellow, green and magenta colors, respectively. The magnetic moments of all magnetically ordered Ru atoms are set to be equal.

In the final refinement, the moments of Ru2, Ru3 and Ru4 were set equal and the moment of Ru1 was fixed to zero. The magnetically ordered Ru atoms reach a moment value $\mu_{exp}$ = 1.97(19) $\mu_B$. Interestingly, similar moment values were obtained for the following double perovskites with nominally pentavalent Ru: $Sr_2YRuO_6$, $\mu_{exp}$ = 1.85(10) $\mu_B$;[75] $Ba_2ScRuO_6$, $\mu_{exp}$ = 2.04(5) $\mu_B$;[76] $Sr_2ScRuO_6$, $\mu_{exp}$ = 1.97(2) $\mu_B$.[76] Assuming a high-spin state for $Ru^{5+}$ with the $4d^3$ electrons occupying the $t_{2g}$ orbitals, a spin-only calculated moment of $\mu_S = g\ S$ = 3.0 $\mu_B$ is expected as discussed above. The significantly lower experimental moment of the $Ru^V$ ion in the magnetically ordered state is possibly systematically reduced by spin-orbital coupling.

**Band structure and calculated magnetic ground state:** The calculated electronic band structure for the rhombohedral $Sr_4NaRu_3O_{12}$ is shown in Figure 9 along high symmetry paths, $\Gamma - M - K - \Gamma - A - L - H - L - K$. A small bandgap of the order of 0.25 eV is observed near the Fermi level which indicates a narrow-gap semiconductor nature. To determine the collinear magnetic ground state, scalar-relativistic pseudopotentials were employed, and the total energy of the system was minimized self-consistently. To understand the spin configuration, we have performed spin-polarized density functional theory (DFT) calculations.

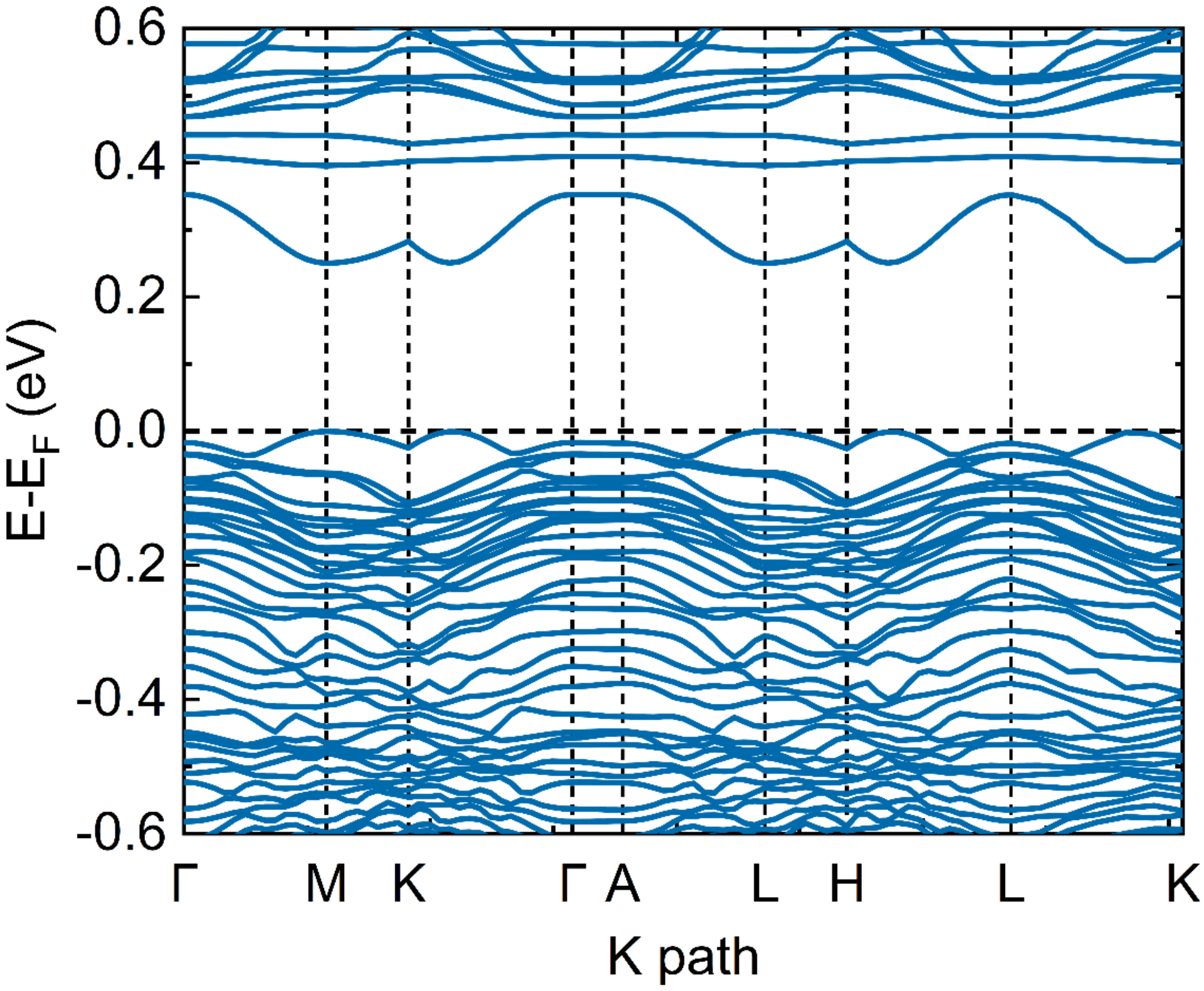

**Figure 9**. Electronic band structure of $Sr_4NaRu_3O_{12}$ calculated with scalar-relativistic pseudopotential in DFT (GGA-PBE) approximation. The Fermi level is indicated by a horizontal dotted line.

The individual magnetic moments, along with their crystal coordinates, are listed in Table S5 in SI. The 4*d* ruthenates are generally dominated by strong Ru 4*d*–O 2*p* hybridization,[77] which leads to the delocalization of the spin-density onto the oxygen ligands and the remaining magnetization goes to the interstitial region.[78] Furthermore, Quantum Espresso calculations account for the spin density within a radius defined by the pseudopotential. As a result, the Ru moments reported in the table should be considered partial, with the remaining contribution on O atoms and in interstitial regions. However, the Ru atoms are considered the main contributors to magnetism in our system. Therefore, we consider the calculated unit cell magnetization in terms of both its total and absolute values to originate from Ru atoms. The absolute magnetization for the unit cell is 87.22 $\mu_B$ for the unit cell containing 36 Ru atoms. This corresponds to an average local moment of approximately 2.42 $\mu_B$ per Ru site, which is in excellent agreement with the expected value for a $Ru^{5+}$ ($d^3$, S = 3/2) state in an octahedral environment. We obtain a negligible total magnetic moment of 0.24 $\mu_B$ for our unit cell with an average of 0.058 $\mu_B$ per Ru atom, affirming that the ground state has a fully compensated antiferromagnetic (AFM) order.

## CONCLUSION

We have synthesized the new ruthenate quadruple perovskites $Sr_4NaRu_3O_{12}$ and $Sr_4LiRu_3O_{12}$ by heating the starting materials in air. $Sr_4NaRu_3O_{12}$ adopts a rare ordered 12*R* quadruple-perovskite structure with predominant Na/Ru *B*-site ordering and exclusively corner-sharing octahedra. It undergoes bulk long-range antiferromagnetic ordering below $T_N$ ~ 270 K as confirmed by susceptibility, heat capacity and neutron diffraction data and shows a magnetic structure with $k$ = (0, 0, 1.5), in which the Ru site at the inversion center remains essentially disordered magnetically. Transport measurements and DFT calculations consistently indicate a narrow-gap semiconducting antiferromagnetic ground state. $Sr_4LiRu_3O_{12}$, by contrast, exhibits a broader anomaly near 110 K and signatures of competing antiferromagnetic and ferromagnetic interactions. The present results establish $Sr_4NaRu_3O_{12}$ as a rare $Ru^V$ oxide combining ordered quadruple-perovskite chemistry with antiferromagnetic ordering close to room temperature.

### Author Contributions

The manuscript was written through the contributions of all authors. All authors have given approval to the final version of the manuscript.

### Conflict of Interest

There are no conflicts of interest.

## ACKNOWLEDGEMENTS

Initial part of this work was partially supported by Würzburg–Dresden Cluster of Excellence '*ctd.qmat'* (project-ID 390858490) under the independent Flex-Fund scheme and later part was supported by ANRF Early Career Research Grant (sanction no.: ANRF/ECRG/2024/001436/CS) during later part. Authors thank Dr. Gudrun Auffermann for the ICP-OES measurements, the central advanced instrumental facility (CAIF, IISER Berhampur) for providing access to XRD, PPMS and XPS facilities.

## Supporting Information

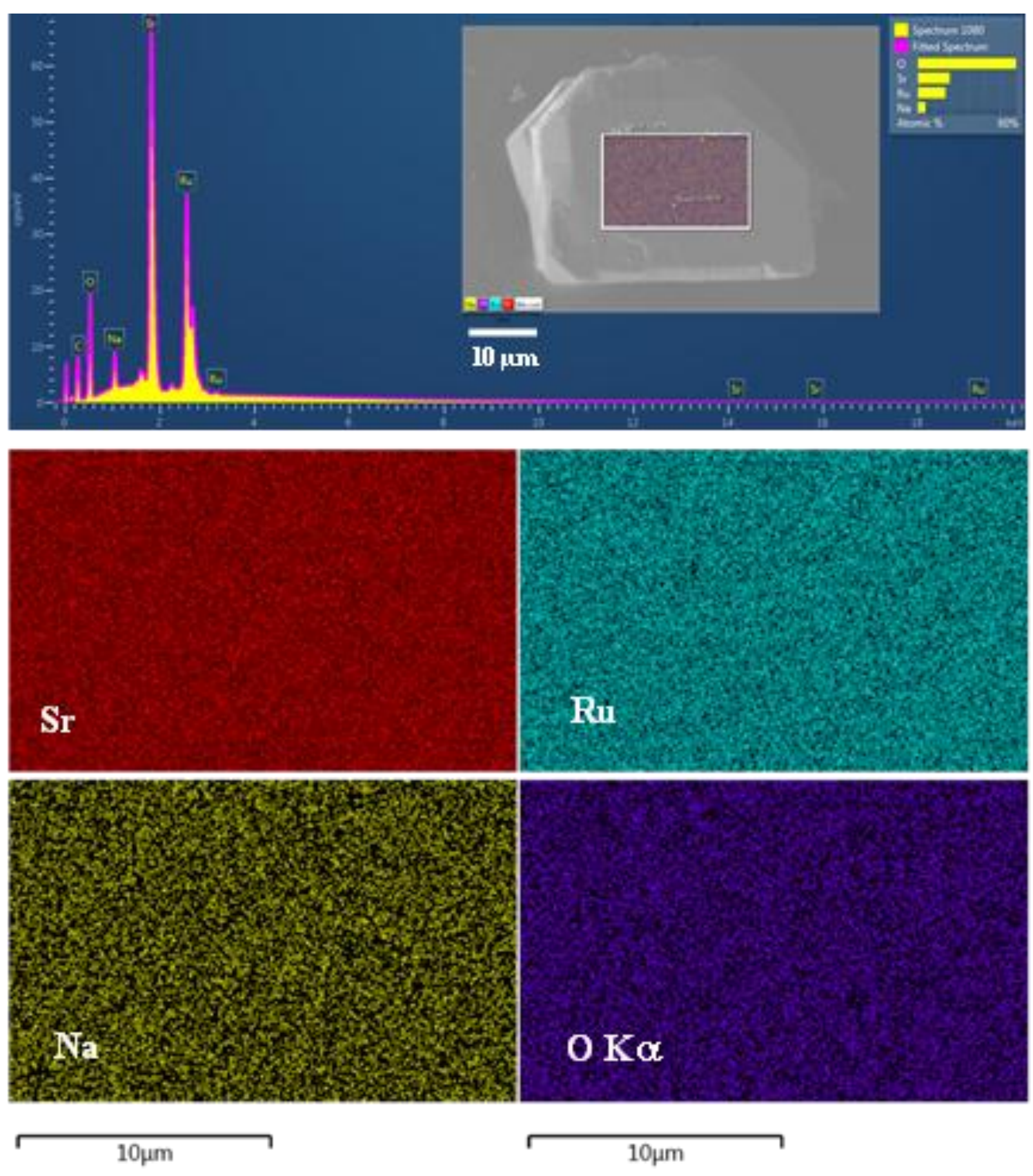

**Figure S1**. SEM spectrum of a $Sr_4NaRu_3O_{12}$ single crystal along with elemental mapping.

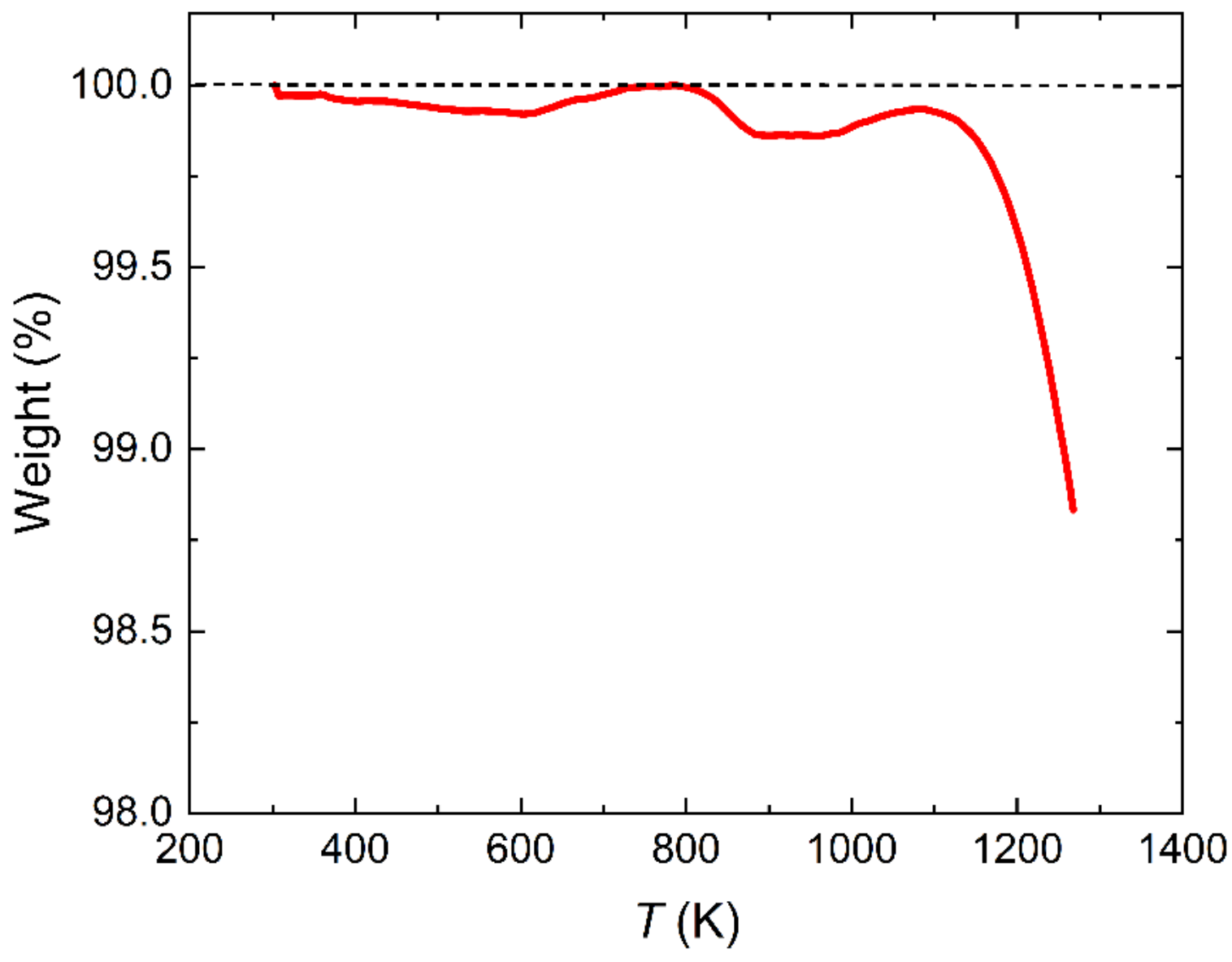

**Figure S2**: Thermal decomposition profile of $Sr_4NaRu_3O_{12}$ under argon atmosphere.

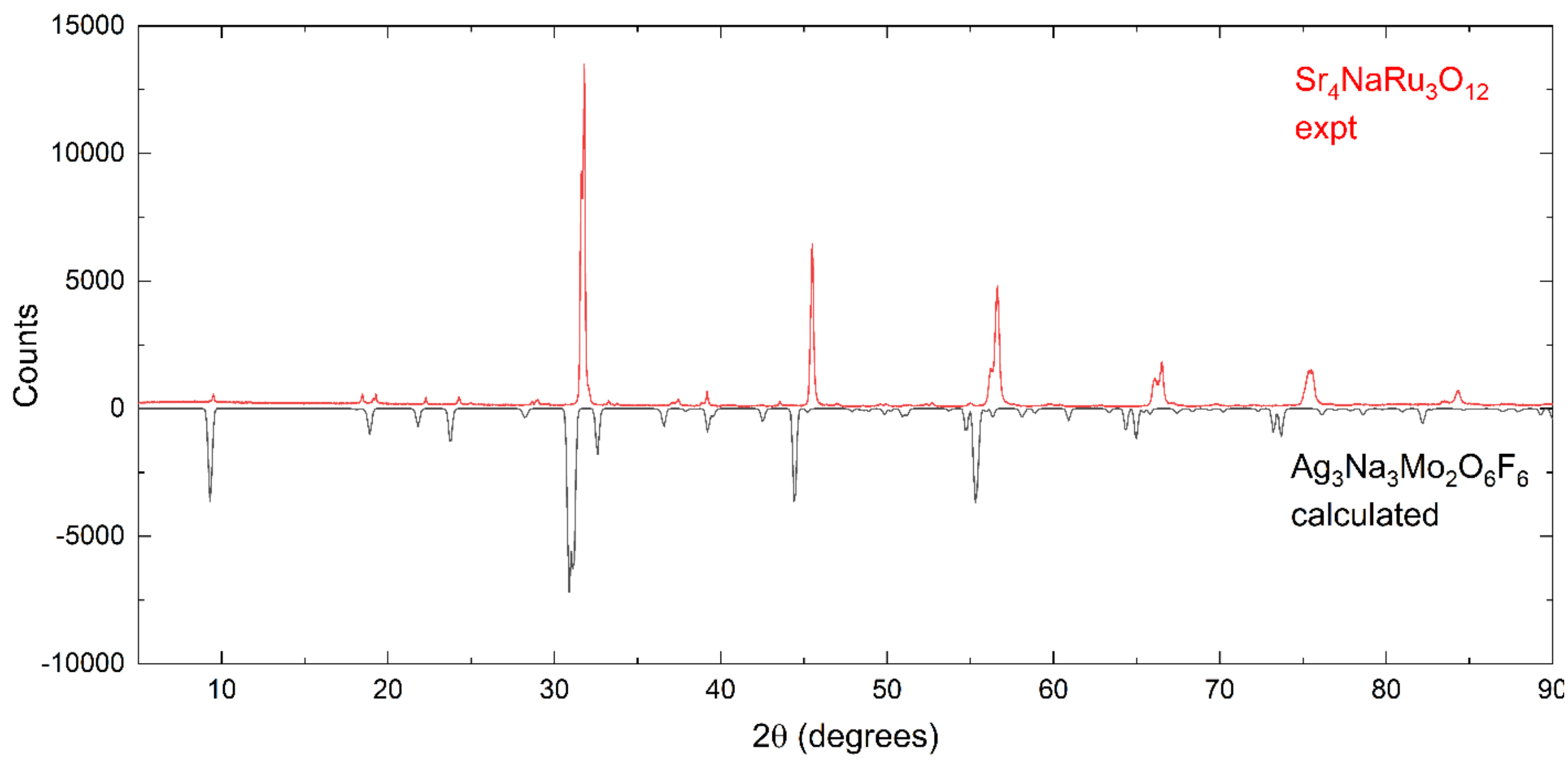

**Figure S3**: Comparison of PXRD patterns of $Sr_4NaRu_3O_{12}$ and $Ag_{1.5}Na_{1.5}MoO_3F_3$. The two patterns are displaced with respect to the *y*-axis due to difference in the *c*-lattice parameters.

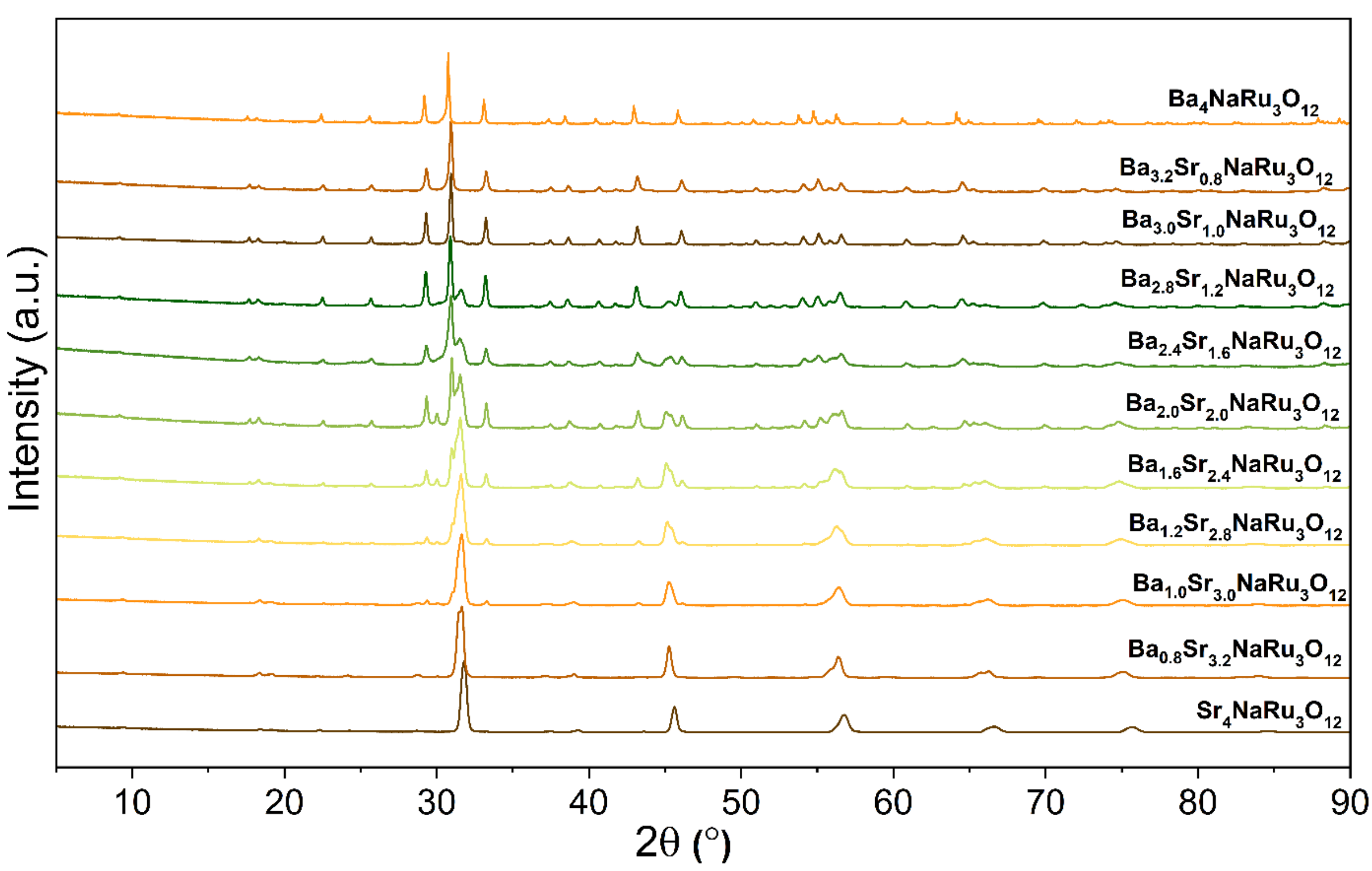

**Figure S4**: PXRD patterns of $(Sr_{1-x}Ba_x)_4NaRu_3O_{12}$ solid solutions.

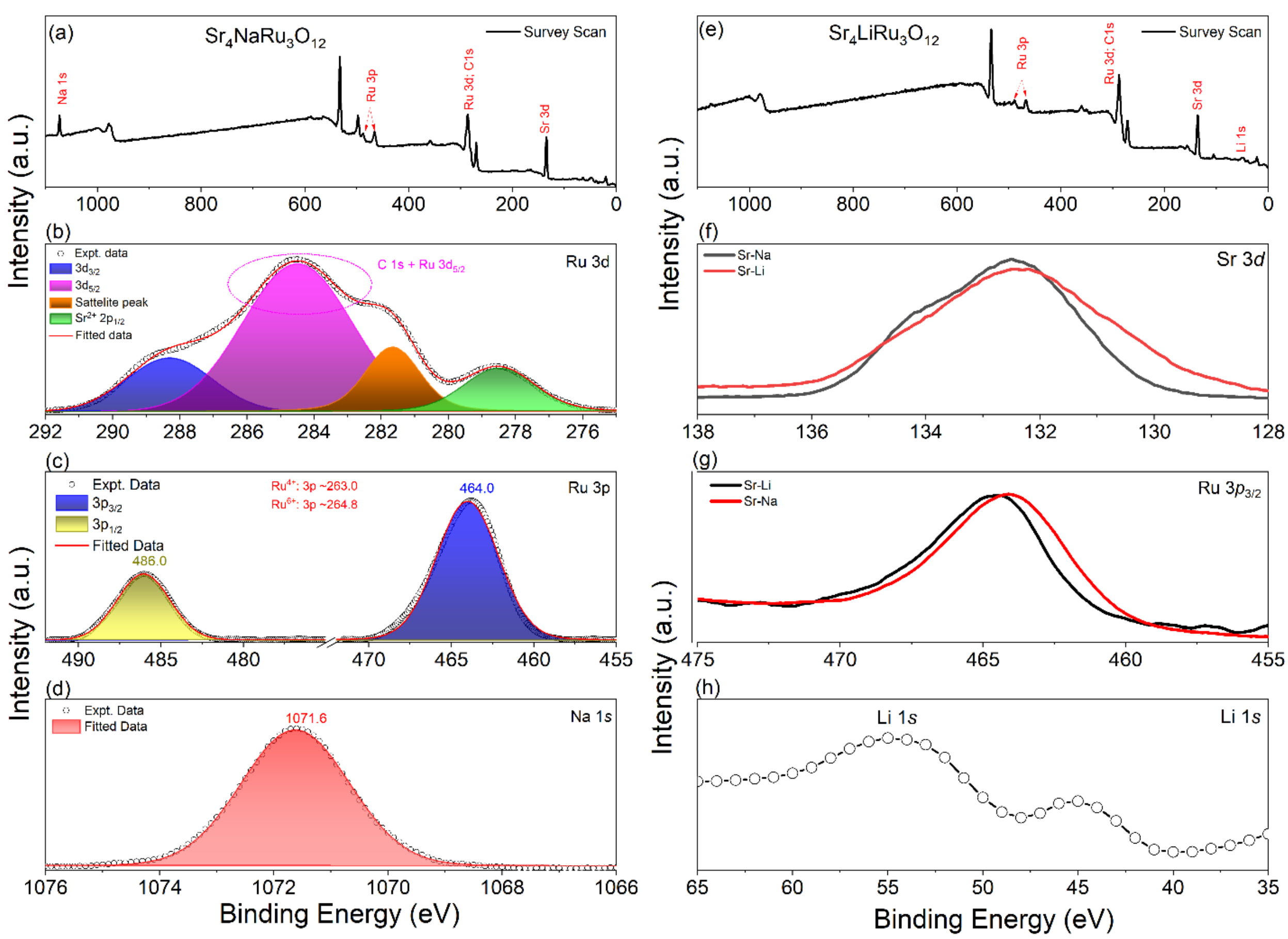

**Figure S5**: (a-d) Fitted core-level XPS spectra of $Sr_4NaRu_3O_{12}$; (e-h) XPS spectra of $Sr_4LiRu_3O_{12}$.

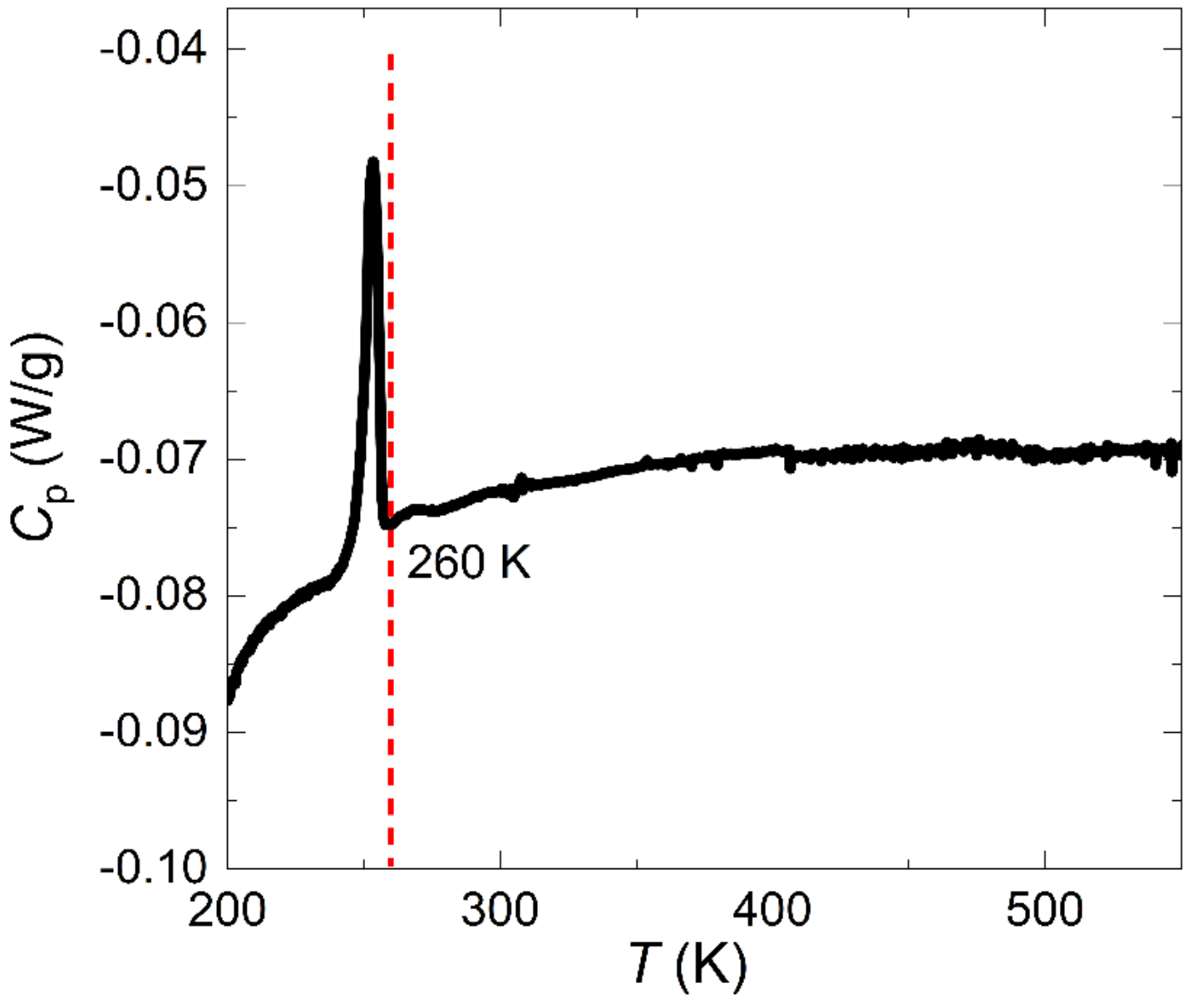

**Figure S6**: DSC curve for $Sr_4NaRu_3O_{12}$ collected during warming run under argon atmosphere.

**Table S1**: Results of the crystal structure refinements of $Sr_4NaRu_3O_{12}$.

| | |
|---|---|
| Temperature | **298(2) K** |
| Empirical formula | $Sr_4NaRu_3O_{12}$ |
| Refined formula | $Sr_4Na_{0.84}Ru_{3.06}O_{12}$ |
| Formula weight, g mol$^{-1}$ | 868.8 |
| Wavelength, Å | 0.71073 |
| Space group (no.) | $R\overline{3}$, (148) |
| Unit cell dimensions, Å | $a$ = 11.2332(5); [$a$ = 11.232(1)]*<br>$c$ = 27.5324(10); [$c$ = 27.727(1)]* |
| Volume, Å$^3$ | 3008.7(2) |
| Density calculated, g cm$^{-3}$ | 5.771 |
| Absorption coefficient mm$^{-1}$ | 25.616 |
| $F$(000) | 4704 |
| $2\theta$ range | 2.22 to 36.17° |
| Index ranges | $-18 \leq h,k \leq 18$<br>$-44 \leq l \leq 44$ |
| Reflections collected | 60588 |
| Independent reflections | 3142 |
| Completeness to θ = 54.81° | 99 % |
| Absorption correction | Muti scan |
| Max. and min. transmission | 0.6567/0.7471 |
| Refinement method | Full-matrix least-squares on $F^2$ |
| Data / restraints / parameter | 3142 / 0 /79 |
| Goodness-of-fit on $F^2$ | 1.6 |
| Final $R$ indices [$I > 3\sigma(I)$] | $R_1$ = 0.0664, w$R_2$ = 0.1635 |
| $R$ indices (all data) | $R_1$ = 0.1548, w$R_2$ = 0.1895 |
| Twin law | 1–10, 0–10, 0 0–1 |
| Twin volume fraction | 0.495(4) |
| Largest diff. peak and hole (e·Å$^{-3}$) | 5.02/–6.24 |

*Values obtained from PXRD data.

**Table S2**: Atomic coordinates for $Sr_4NaRu_3O_{12}$ obtained from single crystal refinement.

| $R\overline{3}$ (148), $a$ = 11.2332(5) Å, $c$ = 27.5324(10) Å: $Z$ = 12 | | | | | | |
|---|---|---|---|---|---|---|
| **Atom** | **Wyck** | **S.O.F.** | ***x/a*** | ***y/b*** | ***z/c*** | **$U_{iso}$ [Å$^2$]** |
| Ru1 | 3a | | 2/3 | 1/3 | 1/3 | |
| Ru2 | 18f | 0.966(4) | 0.8309(3) | 0.66401(7) | 0.4168(1) | |
| Ru3 | 9d | 0.933(4) | 1/2 | 1/2 | 1/2 | |
| Ru4 | 6c | | 0 | 0 | 0.2491(1) | |
| Na1 | 3b | 0.79(1) | 0 | 0 | 1/2 | 0.028(3) |
| Ru5 | 3b | 0.21(1) | 0 | 0 | 1/2 | 0.028(3) |
| Na2 | 9e | 0.852(8) | 1/6 | 5/6 | 1/3 | 0.009(1) |
| Ru6 | 9e | 0.148(8) | 1/6 | 5/6 | 1/3 | 0.009(1) |
| Sr1 | 6c | | 2/3 | 1/3 | 0.4619(1) | |
| Sr2 | 6c | | 1 | 1 | 0.3716(1) | |
| Sr3 | 18f | | 0.6755(1) | 0.8433(2) | 0.4563(1) | |
| Sr4 | 18f | | 0.8416(1) | 0.6660(1) | 0.2914(1) | |
| O1 | 18f | | 0.431(1) | 0.3575(9) | 0.4508(3) | 0.033(2) |
| O2 | 18f | | 1.1022(7) | 0.9478(7) | 0.2868(2) | 0.013(2) |

| O3 | 18f | 0.7409(1) | 0.739(1) | 0.3845(4) | 0.047(3) |
|---|---|---|---|---|---|
| O4 | 18f | 0.912(2) | 0.8183(8) | 0.4581(3) | 0.044(3) |
| O5 | 18f | 0.4443(7) | 0.6129(7) | 0.4594(3) | 0.014(2) |
| O6 | 18f | 0.742(2) | 0.4885(9) | 0.3731(3) | 0.050(3) |
| O7 | 18f | 0.6719(7) | 0.5648(8) | 0.4635(3) | 0.018(2) |
| O8 | 18f | 0.9839(8) | 0.7592(8) | 0.3747(3) | 0.021(2) |

**Table S3**: Selected interatomic distances for $Sr_4NaRu_3O_{12}$ obtained from single crystal refinement.

| Bond | Distance (Å) | Bond | Distance (Å) | Bond | Distance (Å) |
|---|---|---|---|---|---|
| Ru1-O6 | 1.865(9) ×4 | Ru3-O1 | 1.938(9) ×2 | Na1-O4 | 2.111(8) ×6 |
| | 1.86(2) ×2 | Ru3-O7 | 1.965(8) ×2 | Na2-O3 | 2.16(1) ×2 |
| Ru2-O3 | 1.84 (2) | Ru3-O5 | 2.010(9) ×2 | Na2-O8 | 2.120(8) ×2 |
| Ru2-O4 | 1.883(8) | | | Na2-O3 | 2.180(8) ×2 |
| Ru2-O8 | 1.898(7) | Ru4-O2 | 1.847(9) ×3 | | |
| Ru2-O7 | 2.024(7) | Ru4-O5 | 2.001(8) ×3 | | |
| Ru2-O1 | 2.08(2) | | | | |
| Ru2-O6 | 2.089(9) | | | | |

**Table S4**: Anisotropic displacement parameters for $Sr_4NaRu_3O_{12}$. All oxygen and Na atoms were refined isotropically.

| Atom | $U_{11}$ | $U_{22}$ | $U_{33}$ | $U_{12}$ | $U_{13}$ | $U_{23}$ |
|---|---|---|---|---|---|---|
| Ru1 | 0.0090(7) | 0.0090(7) | 0.003(1) | 0.0045(4) | 0.00000 | 0.00000 |
| Ru2 | 0.0052(4) | 0.0074(8) | 0.0051(4) | 0.0028(4) | 0.0012(4) | 0.0006(3) |
| Ru3 | 0.0026(6) | 0.0048(6) | 0.0033(6) | 0.0024(4) | 0.0013(6) | -0.0001(6) |
| Ru4 | 0.0075(7) | 0.0075(7) | 0.0058(8) | 0.0038(4) | 0.00000 | 0.00000 |
| Sr1 | 0.0122(6) | 0.0122(6) | 0.0057(9) | 0.0061(3) | 0.00000 | 0.00000 |
| Sr2 | 0.0153(7) | 0.0153(7) | 0.0081(9) | 0.0076(4) | 0.00000 | 0.00000 |
| Sr3 | 0.0120(4) | 0.0147(5) | 0.0145(5) | 0.0073(5) | 0.0016(3) | 0.0028(3) |
| Sr4 | 0.0120(4) | 0.0147(5) | 0.0145(5) | 0.0073(5) | 0.0016(3) | 0.0028(3) |

**Table S5**: Calculated magnetic moments and atomic coordinates in crystal units of each Ru atom from DFT.

| Atom | *x* (cryst) | *y* (cryst) | *z* (cryst) | Magnetic moment ($\mu_B$/cell) |
|---|---|---|---|---|
| Ru1 | 0.6667 | 0.3333 | 0.3333 | 1.1035 |
| Ru2 | 0.3333 | 0.6667 | 0.6667 | -1.0993 |
| Ru3 | 1.0000 | 0.0000 | 0.0000 | -1.1423 |
| Ru4 | 0.8309 | 0.6640 | 0.4168 | -1.2183 |
| Ru5 | 0.3360 | 0.1669 | 0.4168 | -1.2186 |
| Ru6 | 0.8331 | 0.1691 | 0.4168 | -1.2209 |
| Ru7 | 0.1691 | 0.3360 | 0.5832 | -1.2400 |
| Ru8 | 0.6640 | 0.8331 | 0.5832 | -1.2398 |
| Ru9 | 0.1669 | 0.8309 | 0.5832 | -1.2383 |
| Ru10 | 0.4976 | 0.9973 | 0.7501 | 1.2176 |
| Ru11 | 0.0027 | 0.5002 | 0.7501 | 1.2176 |
| Ru12 | 0.4998 | 0.5024 | 0.7501 | 1.2178 |
| Ru13 | 0.8358 | 0.6693 | 0.9165 | 1.2132 |
| Ru14 | 0.3307 | 0.1664 | 0.9165 | 1.2132 |
| Ru15 | 0.8336 | 0.1642 | 0.9165 | 1.2142 |
| Ru16 | 0.1642 | 0.3307 | 0.0835 | 1.2076 |
| Ru17 | 0.6693 | 0.8336 | 0.0835 | 1.2087 |
| Ru18 | 0.1664 | 0.8358 | 0.0835 | 1.2138 |
| Ru19 | 0.5024 | 0.0027 | 0.2499 | 1.2193 |
| Ru20 | 0.9973 | 0.4998 | 0.2499 | 1.2190 |
| Ru21 | 0.5002 | 0.4976 | 0.2499 | -1.2801 |
| Ru22 | 0.5000 | 0.5000 | 0.5000 | 1.1303 |
| Ru23 | 0.5000 | 0.0000 | 0.5000 | 1.1313 |
| Ru24 | 0.0000 | 0.5000 | 0.5000 | 1.1312 |
| Ru25 | 0.1667 | 0.8333 | 0.8333 | -1.1353 |
| Ru26 | 0.1667 | 0.3333 | 0.8333 | -1.1351 |
| Ru27 | 0.6667 | 0.8333 | 0.8333 | -1.1349 |
| Ru28 | 0.8333 | 0.1667 | 0.1667 | -1.1349 |
| Ru29 | 0.8333 | 0.6667 | 0.1667 | -1.1651 |
| Ru30 | 0.3333 | 0.1667 | 0.1667 | -1.1624 |
| Ru31 | 0.0000 | 0.0000 | 0.2491 | 1.1638 |
| Ru32 | 0.0000 | 0.0000 | 0.7509 | 1.1818 |
| Ru33 | 0.6667 | 0.3333 | 0.5824 | -1.1850 |
| Ru34 | 0.6667 | 0.3333 | 0.0842 | -1.1941 |
| Ru35 | 0.3333 | 0.6667 | 0.9158 | 1.1776 |
| Ru36 | 0.3333 | 0.6667 | 0.4176 | -1.1794 |
| Na | 0.0000 | 0.0000 | 0.5000 | 0.0015 |
| Na | 0.6667 | 0.3333 | 0.8333 | -0.0013 |
| Na | 0.3333 | 0.6667 | 0.1667 | -0.0010 |
| Na | 0.1667 | 0.8333 | 0.3333 | -0.0001 |
| Na | 0.1667 | 0.3334 | 0.3333 | 0.0000 |
| Na | 0.6666 | 0.8333 | 0.3333 | 0.0003 |
| Na | 0.8333 | 0.1667 | 0.6667 | 0.0000 |
| Na | 0.8333 | 0.6666 | 0.6667 | 0.0000 |
| Na | 0.3334 | 0.1667 | 0.6667 | 0.0000 |
| Na | 0.5000 | 0.5000 | 0.0000 | -0.0005 |
| Na | 0.5000 | 0.9999 | 0.0000 | -0.0005 |
| Na | 0.0001 | 0.5000 | 0.0000 | -0.0005 |
| Sr | 0.6667 | 0.3333 | 0.4619 | 0.0084 |
| Sr | 0.3333 | 0.6667 | 0.5381 | 0.0050 |
| Sr | 0.3334 | 0.6666 | 0.7952 | -0.0083 |
| Sr | 1.0000 | 0.0000 | 0.8714 | -0.0083 |
| Sr | 0.0000 | 1.0000 | 0.1286 | -0.0089 |
| Sr | 0.6666 | 0.3334 | 0.2048 | -0.0048 |
| Sr | 0.0000 | 0.0000 | 0.3716 | -0.0097 |
| Sr | 0.0000 | 0.0000 | 0.6284 | -0.0101 |
| Sr | 0.6667 | 0.3333 | 0.7049 | 0.0095 |
| Sr | 0.6667 | 0.3333 | 0.9617 | 0.0100 |
| Sr | 0.3333 | 0.6667 | 0.0383 | 0.0187 |
| Sr | 0.3333 | 0.6667 | 0.2951 | -0.0001 |
| Sr | 0.6755 | 0.8433 | 0.4563 | -0.0078 |
| Sr | 0.1567 | 0.8322 | 0.4563 | -0.0077 |
| Sr | 0.1678 | 0.3245 | 0.4563 | -0.0077 |
| Sr | 0.3245 | 0.1567 | 0.5437 | -0.0079 |
| Sr | 0.8433 | 0.1678 | 0.5437 | -0.0079 |
| Sr | 0.8322 | 0.6755 | 0.5437 | -0.0079 |
| Sr | 0.3422 | 0.1766 | 0.7896 | 0.0076 |
| Sr | 0.8234 | 0.1655 | 0.7896 | 0.0076 |
| Sr | 0.8345 | 0.6578 | 0.7896 | 0.0076 |
| Sr | 0.9912 | 0.4900 | 0.8770 | 0.0077 |
| Sr | 0.5100 | 0.5011 | 0.8770 | 0.0078 |
| Sr | 0.4989 | 0.0088 | 0.8770 | 0.0077 |
| Sr | 0.0088 | 0.5100 | 0.1230 | 0.0049 |
| Sr | 0.4900 | 0.4989 | 0.1230 | 0.0027 |
| Sr | 0.5011 | 0.9912 | 0.1230 | 0.0049 |
| Sr | 0.6578 | 0.8234 | 0.2104 | 0.0031 |
| Sr | 0.1766 | 0.8345 | 0.2104 | 0.0076 |
| Sr | 0.1655 | 0.3422 | 0.2104 | -0.0005 |
| Sr | 0.8416 | 0.6660 | 0.2914 | 0.0016 |
| Sr | 0.3340 | 0.1756 | 0.2914 | 0.0006 |
| Sr | 0.8244 | 0.1584 | 0.2914 | 0.0027 |
| Sr | 0.1584 | 0.3340 | 0.7086 | -0.0001 |
| Sr | 0.6660 | 0.8244 | 0.7086 | -0.0001 |
| Sr | 0.1756 | 0.8416 | 0.7086 | -0.0001 |
| Sr | 0.5083 | 0.9993 | 0.6247 | -0.0031 |
| Sr | 0.0007 | 0.5089 | 0.6247 | -0.0031 |
| Sr | 0.4911 | 0.4917 | 0.6247 | -0.0031 |
| Sr | 0.8251 | 0.6673 | 0.0419 | -0.0030 |
| Sr | 0.3327 | 0.1577 | 0.0419 | -0.0031 |
| Sr | 0.8423 | 0.1749 | 0.0419 | -0.0030 |
| Sr | 0.1749 | 0.3327 | 0.9581 | 0.0032 |
| Sr | 0.6673 | 0.8423 | 0.9581 | 0.0032 |
| Sr | 0.1577 | 0.8251 | 0.9581 | 0.0032 |
| Sr | 0.4917 | 0.0007 | 0.3753 | 0.0001 |
| Sr | 0.9993 | 0.4911 | 0.3753 | 0.0001 |
| Sr | 0.5089 | 0.5083 | 0.3753 | -0.0033 |
| O | 0.4305 | 0.3575 | 0.4508 | 0.0693 |
| O | 0.6425 | 0.0730 | 0.4508 | 0.0689 |
| O | 0.9270 | 0.5695 | 0.4508 | 0.0736 |
| O | 0.5695 | 0.6425 | 0.5492 | 0.0545 |
| O | 0.3575 | 0.9270 | 0.5492 | 0.0550 |
| O | 0.0730 | 0.4305 | 0.5492 | 0.0554 |
| O | 0.0972 | 0.6908 | 0.7841 | -0.0685 |
| O | 0.3092 | 0.4063 | 0.7841 | -0.0686 |
| O | 0.5937 | 0.9028 | 0.7841 | -0.0685 |
| O | 0.2362 | 0.9758 | 0.8825 | -0.0748 |
| O | 0.0242 | 0.2603 | 0.8825 | -0.0746 |
| O | 0.7397 | 0.7638 | 0.8825 | -0.0745 |
| O | 0.7638 | 0.0242 | 0.1175 | -0.0828 |
| O | 0.9758 | 0.7397 | 0.1175 | -0.0814 |
| O | 0.2603 | 0.2362 | 0.1175 | -0.0867 |
| O | 0.9028 | 0.3092 | 0.2159 | -0.0612 |
| O | 0.6908 | 0.5937 | 0.2159 | -0.1825 |
| O | 0.4063 | 0.0972 | 0.2159 | -0.0734 |
| O | 0.1022 | 0.9478 | 0.2868 | 0.1573 |
| O | 0.0522 | 0.1544 | 0.2868 | 0.1534 |

| O | 0.8456 | 0.8978 | 0.2868 | 0.1534 |
|---|---|---|---|---|
| O | 0.8978 | 0.0522 | 0.7132 | 0.1572 |
| O | 0.9478 | 0.8456 | 0.7132 | 0.1571 |
| O | 0.1544 | 0.1022 | 0.7132 | 0.1571 |
| O | 0.7689 | 0.2811 | 0.6201 | -0.1597 |
| O | 0.7189 | 0.4877 | 0.6201 | -0.1598 |
| O | 0.5123 | 0.2311 | 0.6201 | -0.1599 |
| O | 0.5645 | 0.3855 | 0.0465 | -0.1541 |
| O | 0.6145 | 0.1789 | 0.0465 | -0.1522 |
| O | 0.8211 | 0.4355 | 0.0465 | -0.1538 |
| O | 0.4355 | 0.6145 | 0.9535 | 0.1628 |
| O | 0.3855 | 0.8211 | 0.9535 | 0.1619 |
| O | 0.1789 | 0.5645 | 0.9535 | 0.1623 |
| O | 0.2311 | 0.7189 | 0.3799 | -0.1640 |
| O | 0.2811 | 0.5123 | 0.3799 | -0.1572 |
| O | 0.4877 | 0.7689 | 0.3799 | -0.1550 |
| O | 0.7399 | 0.7392 | 0.3845 | -0.1826 |
| O | 0.2608 | 0.0007 | 0.3845 | -0.1803 |
| O | 0.9993 | 0.2601 | 0.3845 | -0.1802 |
| O | 0.2601 | 0.2608 | 0.6155 | -0.1878 |
| O | 0.7392 | 0.9993 | 0.6155 | -0.1876 |
| O | 0.0007 | 0.7399 | 0.6155 | -0.1878 |
| O | 0.4066 | 0.0725 | 0.7178 | 0.1810 |
| O | 0.9275 | 0.3340 | 0.7178 | 0.1811 |
| O | 0.6660 | 0.5934 | 0.7178 | 0.1811 |
| O | 0.9268 | 0.5941 | 0.9488 | 0.1780 |
| O | 0.4059 | 0.3326 | 0.9488 | 0.1782 |
| O | 0.6674 | 0.0732 | 0.9488 | 0.1779 |
| O | 0.0732 | 0.4059 | 0.0512 | 0.1788 |
| O | 0.5941 | 0.6674 | 0.0512 | 0.1779 |
| O | 0.3326 | 0.9268 | 0.0512 | 0.1747 |
| O | 0.5934 | 0.9275 | 0.2822 | 0.1855 |
| O | 0.0725 | 0.6660 | 0.2822 | 0.1871 |
| O | 0.3340 | 0.4066 | 0.2822 | -0.1860 |
| O | 0.9120 | 0.8183 | 0.4581 | -0.1747 |
| O | 0.1817 | 0.0937 | 0.4581 | -0.1751 |
| O | 0.9063 | 0.0880 | 0.4581 | -0.1761 |
| O | 0.0880 | 0.1817 | 0.5419 | -0.1607 |
| O | 0.8183 | 0.9063 | 0.5419 | -0.1608 |
| O | 0.0937 | 0.9120 | 0.5419 | -0.1603 |
| O | 0.5787 | 0.1516 | 0.7914 | 0.1741 |
| O | 0.8484 | 0.4270 | 0.7914 | 0.1741 |
| O | 0.5730 | 0.4213 | 0.7914 | 0.1741 |
| O | 0.7547 | 0.5150 | 0.8752 | 0.1717 |
| O | 0.4850 | 0.2396 | 0.8752 | 0.1718 |
| O | 0.7604 | 0.2453 | 0.8752 | 0.1722 |
| O | 0.2453 | 0.4850 | 0.1248 | 0.1715 |
| O | 0.5150 | 0.7604 | 0.1248 | 0.1711 |
| O | 0.2396 | 0.7547 | 0.1248 | 0.1754 |
| O | 0.4213 | 0.8484 | 0.2086 | 0.1590 |
| O | 0.1516 | 0.5730 | 0.2086 | 0.1584 |
| O | 0.4270 | 0.5787 | 0.2086 | -0.1670 |
| O | 0.4443 | 0.6129 | 0.4594 | 0.0330 |
| O | 0.3871 | 0.8314 | 0.4594 | 0.0332 |
| O | 0.1686 | 0.5557 | 0.4594 | 0.0330 |
| O | 0.5557 | 0.3871 | 0.5406 | 0.0352 |
| O | 0.6129 | 0.1686 | 0.5406 | 0.0349 |
| O | 0.8314 | 0.4443 | 0.5406 | 0.0349 |
| O | 0.1110 | 0.9462 | 0.7927 | -0.0321 |
| O | 0.0538 | 0.1647 | 0.7927 | -0.0322 |
| O | 0.8353 | 0.8890 | 0.7927 | -0.0322 |
| O | 0.2224 | 0.7204 | 0.8739 | -0.0293 |
| O | 0.2796 | 0.5019 | 0.8739 | -0.0294 |
| O | 0.4981 | 0.7776 | 0.8739 | -0.0293 |
| O | 0.7776 | 0.2796 | 0.1261 | -0.1500 |
| O | 0.7204 | 0.4981 | 0.1261 | -0.1500 |
| O | 0.5019 | 0.2224 | 0.1261 | -0.1486 |
| O | 0.8890 | 0.0538 | 0.2073 | -0.0309 |
| O | 0.9462 | 0.8353 | 0.2073 | -0.0334 |
| O | 0.1647 | 0.1110 | 0.2073 | -0.0397 |
| O | 0.7421 | 0.4885 | 0.3731 | 0.0334 |
| O | 0.5115 | 0.2536 | 0.3731 | 0.0338 |
| O | 0.7464 | 0.2579 | 0.3731 | 0.0572 |
| O | 0.2579 | 0.5115 | 0.6269 | -0.1605 |
| O | 0.4885 | 0.7464 | 0.6269 | -0.1605 |
| O | 0.2536 | 0.7421 | 0.6269 | -0.1608 |
| O | 0.4088 | 0.8218 | 0.7064 | -0.0482 |
| O | 0.1782 | 0.5869 | 0.7064 | -0.0481 |
| O | 0.4131 | 0.5912 | 0.7064 | -0.0477 |
| O | 0.9246 | 0.8448 | 0.9602 | -0.0392 |
| O | 0.1552 | 0.0797 | 0.9602 | -0.0397 |
| O | 0.9203 | 0.0754 | 0.9602 | -0.0392 |
| O | 0.0754 | 0.1552 | 0.0398 | -0.0355 |
| O | 0.8448 | 0.9203 | 0.0398 | -0.0363 |
| O | 0.0797 | 0.9246 | 0.0398 | -0.0397 |
| O | 0.5912 | 0.1782 | 0.2936 | 0.1542 |
| O | 0.8218 | 0.4131 | 0.2936 | 0.1533 |
| O | 0.5869 | 0.4088 | 0.2936 | 0.0475 |
| O | 0.6719 | 0.5648 | 0.4635 | 0.0403 |
| O | 0.4352 | 0.1071 | 0.4635 | 0.0439 |
| O | 0.8929 | 0.3281 | 0.4635 | 0.0384 |
| O | 0.3281 | 0.4352 | 0.5365 | 0.0246 |
| O | 0.5648 | 0.8929 | 0.5365 | 0.0252 |
| O | 0.1071 | 0.6719 | 0.5365 | 0.0250 |
| O | 0.3386 | 0.8981 | 0.7968 | -0.0383 |
| O | 0.1019 | 0.4404 | 0.7968 | -0.0384 |
| O | 0.5596 | 0.6614 | 0.7968 | -0.0384 |
| O | 0.9948 | 0.7685 | 0.8698 | -0.0448 |
| O | 0.2315 | 0.2262 | 0.8698 | -0.0448 |
| O | 0.7738 | 0.0052 | 0.8698 | -0.0447 |
| O | 0.0052 | 0.2315 | 0.1302 | -0.0525 |
| O | 0.7685 | 0.7738 | 0.1302 | -0.0535 |
| O | 0.2262 | 0.9948 | 0.1302 | -0.0459 |
| O | 0.6614 | 0.1019 | 0.2032 | -0.0371 |
| O | 0.8981 | 0.5596 | 0.2032 | -0.0455 |
| O | 0.4404 | 0.3386 | 0.2032 | -0.1846 |
| O | 0.9839 | 0.7592 | 0.3747 | -0.1567 |
| O | 0.2408 | 0.2247 | 0.3747 | -0.1593 |
| O | 0.7753 | 0.0161 | 0.3747 | -0.1563 |
| O | 0.0161 | 0.2408 | 0.6253 | -0.1666 |
| O | 0.7592 | 0.7753 | 0.6253 | -0.1668 |
| O | 0.2247 | 0.9839 | 0.6253 | -0.1668 |
| O | 0.6506 | 0.0925 | 0.7080 | 0.1568 |
| O | 0.9075 | 0.5580 | 0.7080 | 0.1567 |
| O | 0.4420 | 0.3494 | 0.7080 | 0.1568 |
| O | 0.6828 | 0.5741 | 0.9586 | 0.1536 |
| O | 0.4259 | 0.1086 | 0.9586 | 0.1532 |
| O | 0.8914 | 0.3172 | 0.9586 | 0.1530 |
| O | 0.3172 | 0.4259 | 0.0414 | 0.1555 |
| O | 0.5741 | 0.8914 | 0.0414 | 0.1572 |
| O | 0.1086 | 0.6828 | 0.0414 | 0.1547 |
| O | 0.3494 | 0.9075 | 0.2920 | 0.1623 |
| O | 0.0925 | 0.4420 | 0.2920 | 0.1598 |
| O | 0.5580 | 0.6506 | 0.2920 | -0.1570 |